\documentclass[aps,showpacs,superscriptaddress,preprintnumbers,nofootinbib]{revtex4}
\usepackage{slashed,color,amsmath,amssymb,mathrsfs}
\usepackage{graphicx}
\usepackage{hyperref}
\usepackage{longtable}
\usepackage{array}

\allowdisplaybreaks
 \def\sk{\Sigma_k}
 \def\skp{\Sigma_k'}
 \def\skpp{\Sigma_k''}
 \def\skppp{\Sigma_k'''}

 \def\la{\langle}
 \def\ra{\rangle}

\begin{document}
\preprint{}
\title{Computation of the $O(p^6)$ order low-energy constants: an update}
\author{Shao-Zhou Jiang}
\email[]{jsz@gxu.edu.cn}
\affiliation{Department of Physics and GXU-NAOC Center for Astrophysics and Space Sciences, Guangxi University,
Nanning, Guangxi 530004, P.R.China}
\affiliation{Guangxi Key Laboratory for the Relativistic Astrophysics, Nanning 530004, P.R.China }
\author{Zhen-Long Wei}
\email[]{weizhenlong@mail.gxu.cn}
\author{Qing-Sen Chen}
\email[]{chenqs@mail.gxu.cn}
\affiliation{Department of Physics and GXU-NAOC Center for Astrophysics and Space Sciences, Guangxi University,
Nanning, Guangxi 530004, P.R.China}

\author{Qing Wang}
\email[Corresponding author.]{wangq@mail.tsinghua.edu.cn}
\affiliation{Department of Physics, Tsinghua University, Beijing 100084, P.R.China}
\affiliation{Center for High Energy Physics, Tsinghua University, Beijing 100084, P.R.China}
\affiliation{Collaborative Innovation Center of Quantum Matter, Beijing 100084, P.R.China}

 \begin{abstract}

 We update our original low-energy constants to the $O(p^6)$ order, including two and three flavours,
 the normal and anomalous ones. Following a comparative analysis, the $O(p^4)$ results are considered better.
 In the $O(p^6)$ order, most of our results are consistent or better with those we have found in the literature,
 although several are worse.
 \end{abstract}
  \maketitle
\section{Introduction}
 Chiral perturbation theory(ChPT) is a useful method to solve low-energy mesons interactions.
 A significant problem is to establish the chiral Lagrangian (CL).
 Until now, the monomials of CL have been obtained to the $O(p^6)$ order \cite{weinberg,GS1,GS2,p61,p62,p6p,p6a1,p6a2,tensor1,U3,ourf},
 including both the normal and anomalous parts, two and three flavours, and the special unitary and unitary groups.
 The $O(p^6)$ order monomials seem accurate enough to describe the present experiments.
 In constrast, the coefficients for each of the monomials, called the low-energy constants (LECs),
 also play an important role of ChPT.
 Concrete numerical results need them.
 However ChPT does not provide these LECs, and hence they need to be obtained from other theories.
 Because of non-perturbative effects, exact values for the LECs are hard to obtain, especially at the $O(p^6)$ order.
 At the $O(p^4)$ order, although exact results have not been obtained, values from different methods are close.
 Their signs and magnitudes are the same.
 At the $O(p^6)$ order, there are various methods to obtain LECs, the more common ones being
 resonance chiral theory \cite{tpf,res2}, sum rules \cite{sumrules}, lattice QCD \cite{lattr,latt}
 holographic theory \cite{Colangelo:2012ipa}, QCD \cite{our5}, and global fit \cite{gfit1}.
 Of course, LECs can also be extracted from experimental data.
 Some methods can obtain a value for a single LEC; while others produce values for combinations of LECs.
 Each method has its advantages and disadvantages, but some values do come with large errors because of non-perturbative QCD processes.

 Several years ago, values of a few LECs obtained by different methods were scattered in various references.
 A large number of LECs remained unknown in value. To taking advantage of ChPT, LECs are needed for numerical calculations.
 Our motivation was to explore a method to obtain all the LECs in a single calculation.
 We hoped the method would be model-independent, as had been developed from the underlying QCD theory.
 Nevertheless at that time, the theoretical analysis was poor, and we only used simple methods that involved some rough approximations \cite{our5,oura2}.
 Although expedient, preliminary results at least were obtained.

 Given that the approximations are too rough, the results did yield a gauge-invariant, nonlocal, dynamical-quark (GND model) model \cite{GND}.
 Ref. \cite{gfit1} details checks of our $O(p^6)$  LECs ($C_i$) via a global fit of the $O(p^4)$ LECs ($L_i$) to the $O(p^6)$ order.
 The $\chi^2$ divided by the degrees of freedom ($\chi^2/$dof) is $4.13/4$.
 If all LECs are multiplied by 0.27, $\chi^2/$dof=1.20/3 and hence the $C_i$ values appeared to be too large.
 In statistical theory, when dof is large enough (a typical choice is dof$>$30), $\chi^2/$dof$\sim1$.
 However at only 4, the dof is too small and it is not sufficiently large to assess the reliability of the calculation by $\chi^2/$dof.
 Because of the $\chi^2$ distribution, $\chi^2/$dof is also not as small as it should be.
 Furthermore the $O(p^4)$ and $O(p^6)$ LECs are independent.
 Evaluating the confidence level of the $C_i$ by $L_i$ is not suitable, especially when dof is small.
 A more believable assessment can be garnered by comparing experiment data.
 However the absolute values of our $L_i$ are too large, and may include some systematic errors in the calculation,
 such as those associated with the large $N_C$ limit.
 These systematic errors must exist at all orders.
 With the passing of time, it is now the moment improve the $O(p^6)$ LECs.
 While tedious, these needs to be improved step by step.
 This paper research analyses the origin of the systematic errors and tries to remedy them as precisely as possible.

 This paper is organized as follows: In Sec. \ref{rev}, we review our method to obtain the CL from QCD
 and introduce a more precise approximation. In Sec. \ref{cat}, a concrete method to calculate LECs is introduced.
 In Sec. \ref{results}, we list our results for the LECs, both normal and anomalous, up to and including order $O(p^6)$.
 In Sec. \ref{comp}, we compare our results with others in the literature, and check for new predictions.
 Section \ref{summ} concludes with a summary.

\section{Review and Improvements over Previous Calculations of LECs}\label{rev}

 In a previous work, we obtained the action of the chiral theory in the large $N_c$ limit.
 Drived from first principles QCD, it takes the form \cite{WQ0,WQ1}
 \begin{eqnarray}
 S_{\rm eff}
 &=&-iN_c{\rm Tr}\ln[i\partial\!\!\!/+J_{\Omega}-\Pi_{\Omega c}]
 +N_c\int d^{4}xd^{4}x'
 \Phi^{\sigma\rho}_{\Omega c}(x,x')\Pi^{\sigma\rho}_{\Omega c}(x,x')
 +N_c\sum^{\infty}_{n=2}{\int}d^{4}x_1\cdots d^4x_{n}'\nonumber\\
 &&\times
 \frac{(-i)^{n}(N_c g_s^2)^{n-1}}{n!}\bar{G}^{\sigma_1\cdots\sigma_n}_{\rho_1
 \cdots\rho_n}(x_1,x'_1,\cdots,x_n,x'_n)
 \Phi^{\sigma_1\rho_1}_{\Omega c}(x_1 ,x'_1)\cdots
 \Phi^{\sigma_n\rho_n}_{\Omega c}(x_n ,x'_n)\nonumber\\
 &&+iN_c \int d^4x~{\rm tr}_{lf}\{\Xi_c(x)[-i\sin\frac{\vartheta_c(x)}{N_f}
 +\gamma_5\cos\frac{\vartheta_c(x)}{N_f}]\Phi_{\Omega,c}^T(x,x)\},
 \label{Snorm}
 \end{eqnarray}
 in which $J_{\Omega}$ is the external source $J$  including currents and densities
 after making a Goldstone-field-dependent chiral rotation $\Omega$:
 \begin{eqnarray}
 J_{\Omega}=[\Omega P_R+\Omega^{\dag}P_L][J+i\slashed{\partial}][\Omega P_R+\Omega^\dag P_L]
 =\slashed{v}_\Omega+\slashed{a}_\Omega\gamma_5-s_\Omega+ip_\Omega\gamma_5,\hspace{0.3cm}
 J=\slashed{v}+\slashed{a}\gamma_5-s+ip\gamma_5,\hspace{0.3cm}U=\Omega^2\;.\label{JOmega}
 \end{eqnarray}
 where $\Phi_{\Omega c}$ and $\Pi_{\Omega c}$ are, respectively, the two-point rotated-quark
 Green's functions and the interaction part of two-point rotated-quark
 vertices in the presence of external sources; $\Phi_{\Omega c}$ is defined by
 \begin{eqnarray}
 &&\Phi_{\Omega c}^{\sigma\rho}(x,y)\equiv
 \frac{1}{N_c}\langle\overline{\psi}_{\Omega}^{\sigma}(x)\psi_{\Omega}^{\rho}(y)\rangle
 =-i[(i\slashed{\partial}+J_{\Omega}-\Pi_{\Omega c})^{-1}]^{\rho\sigma}(y,x),\hspace{0.3cm}
 \psi_{\Omega}(x)\equiv[\Omega(x)P_L+\Omega^\dag(x)P_R]\psi(x),\label{PhiPi}\\
 &&\Pi_{\Omega c}^{\sigma\rho}(x,y)=-\tilde{\Xi}^{\sigma\rho}(x)\delta^4(x-y)
 -\sum^{\infty}_{n=1}{\int}d^{4}x_1\cdots{d^4}x_{n}
 d^{4}x_{1}'\cdots{d^4}x_{n}'\frac{(-i)^{n+1}(N_c g_s^2)^n}{n!}\nonumber\\
 &&\hspace{0.2cm}\times\overline{G}^{\sigma\sigma_1\cdots\sigma_n}_{\rho\rho_1
 \cdots\rho_n}(x,y,x_1,x'_1,\cdots,x_n,x'_n)
 \Phi_{\Omega c}^{\sigma_1\rho_1}(x_1 ,x'_1)\cdots
 \Phi_{\Omega c}^{\sigma_n\rho_n}(x_n ,x'_n),\label{fineqNc}
 \end{eqnarray}
 with subscript $_c$ denoting the classical field and $\psi(x)$ the
 light quark fields. $\bar{G}^{\sigma_1\cdots\sigma_n}_{\rho_1
 \cdots\rho_n}(x_1,x'_1,\cdots,x_n,x'_n)$ is the effective gluon $n$-point
 Green's function and $g_s$ is the coupling constant of QCD.
 The $\Phi_{\Omega c}$ and $\Pi_{\Omega c}$ are related by the first equation of \eqref{PhiPi} and determined by
 \begin{eqnarray}
 &&[\Phi_{\Omega c}+\tilde{\Xi}]^{\sigma\rho}+\sum^{\infty}_{n=1}{\int}d^4x_1d^4x'_1\cdots{d^4}x_n
 d^4x'_n\frac{(-i)^{n+1}(N_c g_s^2)^n}{n!}\overline{G}^{\sigma\sigma_1\cdots\sigma_n}_{\rho\rho_1
 \cdots\rho_n}(x,y,x_1,x'_1,\cdots,x_n,x'_n)\nonumber\\
 &&\times \Phi_{\Omega c}^{\sigma_1\rho_1}(x_1 ,x'_1)\cdots
 \Phi_{\Omega c}^{\sigma_n\rho_n}(x_n,x'_n)=O(\frac{1}{N_c}),\label{SDE}
 \end{eqnarray}
 where $\vartheta_c$ is the phase angle of the $U(1)$ factor,
 and $\Xi_c$ and $\tilde{\Xi}$ are two parameters in the calculation, defined by Eqs. (21) and (65) in Ref.\cite{WQ0}.
 In this work, they have little importance and are neglected.
 Eq. (\ref{SDE}) is the Schwinger-Dyson equation (SDE) in the presence of the rotated external source.
 In Ref.\cite{WQ1}, we have assumed an approximate solution of (\ref{SDE}) given by
 \begin{eqnarray}
 \Pi^{\sigma\rho}_{\Omega c}(x,y)=[\Sigma(\overline{\nabla}^2_x)]^{\sigma\rho}\delta^4(x-y),\hspace{3cm}
 \overline{\nabla}^{\mu}_x=\partial^{\mu}_x-iv_{\Omega}^{\mu}(x)\;,\label{defsig}
 \end{eqnarray}
 where $\Sigma$ is the quark self-energy which satisfies SDE (\ref{SDE}) with vanishing rotated external source.
 Under the ladder approximation, this SDE in Euclidean space-time is reduced to the standard form
 \begin{eqnarray}
 \Sigma(p^2)-3C_2(R)\int\frac{d^4q}{4\pi^3}\frac{\alpha_s[(p-q)^2]}{(p-q)^2}
 \frac{\Sigma(q^2)}{q^2+\Sigma^2(q^2)}=0\;, \label{eq0}
 \end{eqnarray}
 where $\alpha_s(p^2)$ is the running coupling constant of QCD which depends on $N_C$ and the number of quark flavours,
 and $C_2(R)$ is the second-order Casimir operator of the quark representation $R$.
 In this work, the quarks belonging to the $SU(N_C)$ fundamental representation,
 and therefore $C_2(R)=(N_c^2-1)/2N_c$; in the large $N_C$ limit, the second term is neglected.

 In our previous article, because of the computational complexity, we did not calculate all terms in Eq. \eqref{Snorm},
 but introduced some approximations to reduced the last three terms and leave only the first.
 The approximations are rough, and the results reproduce those of the GND model \cite{GND}.
 The last three terms could have introduced some systematic errors.
 The absolute values of the previous results seem too large when compared with others in the literature.
 The purpose of this paper is to remedy these errors partially
 and produce a more credible result.

 For now, we have no idea how to solve the last term in \eqref{Snorm} and hence omit it,;
 we shall only focus on the second and third terms.
 The infinite sum in both the third terms in \eqref{Snorm} and the second term in \eqref{fineqNc}
 appear similar but they are not equal.
 To find the source of the systematic errors and to maintain a manageable calculation
 we only retain the two-point contributions,
 the terms with $n=2$ in \eqref{Snorm} and $n=1$ in \eqref{fineqNc}.
 When we simplify SDE from \eqref{SDE} to \eqref{eq0}, we also include $n=1$ in \eqref{SDE}.
 The present approximation and the approximation for SDE are all to first order,
 omitting all higher order contributions; they are then at the same level of accuracy.
 The other terms describe mesonics interactions, which we consider as less important when compared previous terms.
 The final results will substantiate this decision. Hence we shall add an additional effective action,
 \begin{eqnarray}
 \Delta S_{\rm eff}&\sim&\frac{1}{2}N_c\int d^{4}xd^{4}x'
 \Phi^{\sigma\rho}_{\Omega c}(x,x')\Pi^{\sigma\rho}_{\Omega c}(x,x')\label{alag0}\\
 &\sim&-\frac{i}{2}N_c\mathrm{Tr}
 [(i\slashed{\partial}+J_{\Omega}
 -\Sigma(\overline{\nabla}^2))^{-1}\Sigma(\overline{\nabla}^2)].\label{alag}
 \end{eqnarray}
 the approximation of which has been used in \eqref{defsig}.

 With the same considerations for the anomalous parts, we only calculated the first term in \eqref{Snorm},
 although using a different method to introduce the fifth dimensional integral \cite{oura2}.
 We need not repeat this as the additional effective Lagrangian is the same as \eqref{alag}.
 Therefore, for the present study, we calculate \eqref{alag} including both the normal and anomalous parts.

 \section{Calculation of the Additional Terms}\label{cat}

 To calculate the additional terms, we first use the Wick rotation to change \eqref{alag} to Euclidean spacetime as \cite{our5,oura2},
 and then expand it as a Taylor series,
 \begin{eqnarray}
 \Delta S_{\rm eff}
 &=&-\frac{1}{2}N_c\mathrm{Tr}
 [(D+\Sigma(-\overline{\nabla}^2))^{-1}\Sigma(-\overline{\nabla}^2)]\\
 &=&-\frac{1}{2}N_c\mathrm{tr}
 [(-i\slashed{k}+D+\Sigma((k+i\overline{\nabla})^2))^{-1}\Sigma((k+i\overline{\nabla})^2)]\\
 &=&-\frac{1}{2}N_c\int d^4x\int\frac{d^4k}{(2\pi)^4}\mathrm{tr}[(\sk X+i\slashed{k}X)\sum_{n=0}^{\infty}(-1)^n[(D+\Sigma_{1})(\sk X+i\slashed{k}X)]^n
 \Sigma((k+i\overline{\nabla})^2)],\label{dels}
 \end{eqnarray}
 where ``Tr" includes the traces over coordinate space, spinor space and flavour space,
 ``tr" includes only the traces over spinor and flavour spacec,
 $D\equiv\overline{\slashed{\nabla}}-i\slashed{a}_\Omega\gamma_5-s_\Omega+ip_\Omega\gamma_5$,
 and $\Sigma_1\equiv\Sigma((k+i\overline{\nabla})^2)-\Sigma(k^2)$.

 To a given order ($O(p^2)$, $O(p^4)$ or $O(p^6)$), after the tracing over the spinor space, the series are
 \begin{eqnarray}
 \Delta S_{\rm eff}^{2n}=\int d^4x\sum_{k=1}^m a_k \la O_{\Omega,k}\ra \label{soomega}
 \end{eqnarray}
 where $a_k$ are coefficients, $O_{\Omega,k}$ are monomials with flavour indices including
 $\overline{\nabla}^\mu,a_\Omega^\mu,s_\Omega$ $p_\Omega$, and ``$\la\ldots\ra$" means trace over flavour space.
 In \eqref{soomega}, we have used the basic relations to simplify the results,
 including trace relations, the Einstein summation convention, and for the anomalous terms also including the Schouten identity.
 Nevertheless, not all of the $O_{\Omega,k}$ are independent.

 Generally, the number of $O_{\Omega,k}$ is larger than the number of linear independent terms $O_l$.
 Specifically, the relationship between the two is given by
 \begin{eqnarray}
 \la O_l\ra=\sum_{k=1}^m A_{lk}\la O_{\Omega,k}\ra,\hspace{0.2cm}l=1,2,3,\cdots,M\label{lrs0}
 \end{eqnarray}
 where $M$ is the number of LECs up to a given order.\footnote{The building blocks of $O_l$ and $O_{\Omega}$ are different.
 Their relations can be found in Table XV in Ref.\cite{our5}}
 This implies that generally $A_{lk}$ is not a square matrix.
 The reduced row echelon form of $A_{lk}$ is
 \begin{eqnarray}
 A_{lk}\to B_{lk}=
 \left(\begin{array}{cccccccccc}
 1 & C_{12} &  O & C_{14} & O & \cdots & \cdots & \cdots \\
   &   &  1 & C_{24} & O & \cdots & \cdots & \cdots \\
   &   &    & \cdots & \cdots & \cdots & \cdots & \cdots \\
   &   &    &   & 1 & \cdots & \cdots & \cdots
 \end{array}\right),\label{lrs1}
 \end{eqnarray}
 where the bottom-left corner contains just zero elements wiht $O$ representing a zero matrix of the appropriate dimension
 $C$ representing possibly a non-zero matrix and ``$\cdots$" also nonzero matrices.
 The rank of $A_{lk}$ or $B_{lk}$ is equal to the number of independent linear bases
 with each nonzero row-vector in $B_{lk}$ corresponding to a linear basis in $\la O_{\Omega}\ra$.
 We select those $O_{\Omega,k'}$ that are independent,
 and set $B_{k'k'}=1$ and $B_{k'k'}$ to be the first non-zero elements in the $k'$th row in $B_{lk}$.
 All dependent $O_{\Omega,k}$ can be replaced by $O_{\Omega,k'}$.

 Without using the Cayley-Hamilton relations, the LECs $K_l$ for arbitrary $N_f$ flavours are defined as
 \begin{eqnarray}
 S_{\rm eff}^{2n}=\int d^4x\sum_{l=1}^M K_l \la O_l\ra=\int d^4x\sum_{l=1}^M\sum_{k=1}^m K_l A_{lk}\la O_{\Omega,k}\ra.\label{dlecs}
 \end{eqnarray}
 In the second equation, we have has used \eqref{lrs0}.
 Comparing \eqref{soomega} and \eqref{dlecs}, because all the relations in $N_f$ flavours have been used
 the coefficients in front of $\la O_{\Omega,k}\ra$ need to be equal.
 \begin{eqnarray}
 a_k=\sum_{l=1}^M K_l A_{lk},\hspace{0.3cm}k=1,2,3,\ldots m.\label{rkk}
 \end{eqnarray}
 In \eqref{rkk}, there are $m$ equations in $M$ variables with $m>M$.
 We select the $M$ independent $O_{\Omega,k'}$ to solve the equations,
 leaving the remaining $m-M$ equations as constraints. Hence the additional LECs are
 \begin{eqnarray}
 \Delta K_{l}=\sum_{k'} a_{k'}A^{-1}_{k'l}.
 \end{eqnarray}
 Replacing $\Delta K_{l}$ in \eqref{rkk}, the second equal sign must hold.
 These the are the constraint equations.

 Finally, using the Cayley-Hamilton relations, all the LECs can be obtained for three and two flavours.

 \section{Results}\label{results}
 \subsection{Order $O(p^2)$ and $O(p^4)$}
 To the $O(p^2)$ order, the additional analytical LECs are
 \begin{eqnarray}
 \Delta F_0^2&=&\int\frac{d^4k}{(2\pi)^4}[
 4 \sk^2 X^2 -6 \sk^4 X^3],\notag\\
 \Delta F_0^2B_0&=&\int\frac{d^4k}{(2\pi)^4}[2 \sk X
 -4 \sk^3 X^2 ],\notag\\
 X&\equiv&\frac{1}{k^2+\sk^2},\hspace{0.3cm}\sk\equiv\Sigma(k^2).\label{ap2}
 \end{eqnarray}
 The analytical results in \eqref{ap2} and the higher order in the following are all set in Euclidean spacetime.
 A more precise result, with all terms on the right hand side of \eqref{Snorm}, has been given in \cite{WQ0}.
 This is just the Pagels-Stokar formula for $F_0^2$
 and also the $\Lambda\to\infty$ results in our previous work \cite{our5}.
 It seems that only the first term in \eqref{Snorm} is sufficient to give the $O(p^2)$ results,
 the other terms do not contribute at the $O(p^2)$ order.
 In considering the accuracy, we use the original results in \cite{our5} at the $O(p^2)$ order.
 The method described in \cite{WQ0}, however, is not easily extendible to higher order,
 and there is no proof that indicates that the other terms in \eqref{Snorm} make no contribution to the LECs.
 Hence at higher order, we consider \eqref{dels}.

 At the $O(p^4)$ order, the additional analytical results are
 \begin{eqnarray}
 \Delta C_{2}&=&\int\frac{d^4k}{(2\pi)^4}\bigg[- \skp^2 X
 +11 \sk^2\skp^2 X^2
 -34 \sk^4\skp^2 X^3
 -2 \sk^4 X^4
 +40 \sk^6\skp^2 X^4
 +4 \sk^6 X^5
 -16 \sk^8\skp^2 X^5
 \bigg],\notag\\
 \Delta C_{3}&=&\int\frac{d^4k}{(2\pi)^4}\bigg[-\frac{1}{2} \skp^2 X
 +\frac{49}{6} \sk^2\skp^2 X^2
 -\frac{2}{3} \sk^2 X^3
 -\frac{83}{3} \sk^4\skp^2 X^3
 -\frac{5}{3} \sk^4 X^4
 +\frac{100}{3} \sk^6\skp^2 X^4\notag\\
 &&+\frac{10}{3} \sk^6 X^5
 -\frac{40}{3} \sk^8\skp^2 X^5\bigg],\notag\\
 \Delta C_{4}&=&\int\frac{d^4k}{(2\pi)^4}\bigg[-\frac{26}{3} \sk^2 X^3
 +\frac{118}{3} \sk^4 X^4
 -\frac{104}{3} \sk^6 X^5
 \bigg],\notag\\
 \Delta C_{5}&=&\int\frac{d^4k}{(2\pi)^4}\bigg[\frac{14}{3} \sk^2 X^3
 -\frac{52}{3} \sk^4 X^4
 +\frac{44}{3} \sk^6 X^5
 \bigg],\notag\\
 \Delta C_{6}&=&\int\frac{d^4k}{(2\pi)^4}\bigg[6 \sk^2 X^2
 -8 \sk^4 X^3
 \bigg],\notag\\
 \Delta C_{7}&=&\int\frac{d^4k}{(2\pi)^4}\bigg[2 \sk^2 X^2
 \bigg],\notag\\
 \Delta C_{8}&=&\int\frac{d^4k}{(2\pi)^4}\bigg[-4 \sk X^2
 +34 \sk^3 X^3
 -36 \sk^5 X^4
 \bigg],\notag\\
 \Delta C_{9}&=&\int\frac{d^4k}{(2\pi)^4}\bigg[ \skp^2 X
 -\frac{2}{3} \sk\skp X^2
 -3 \sk^2\skp^2 X^2
 -\frac{2}{3} \sk^2 X^3
 +\frac{2}{3} \sk^3\skp X^3
 +2 \sk^4\skp^2 X^3
 \bigg],\notag\\
 \Delta C_{10}&=&\int\frac{d^4k}{(2\pi)^4}\bigg[-2i \skp^2 X
 +\frac{82i}{3} \sk^2\skp^2 X^2
 +\frac{20i}{3} \sk^2 X^3
 -\frac{268i}{3} \sk^4\skp^2 X^3
 -\frac{52i}{3} \sk^4 X^4
 +\frac{320i}{3} \sk^6\skp^2 X^4\notag\\
 &&+\frac{32i}{3} \sk^6 X^5
 -\frac{128i}{3} \sk^8\skp^2 X^5\bigg],\notag\\
 \Delta C_{11}&=&\int\frac{d^4k}{(2\pi)^4}\bigg[-2 \sk X^2
 +6 \sk^3 X^3
 \bigg].
 \end{eqnarray}
 The definitions of $C_i$ can be found in Eq. (19) of \cite{GND},
 and the relations between $C_i$ and common $L_i$ can be found in Eq. (24) in \cite{GND}.

 To obtain numerical results, we use the same quark self-energy $\sk$ as in \cite{WQ1,our5}
 with the running coupling constant $\alpha_s(p^2)$ of \cite{alphas}.
 To complete the integral, two other input parameters are needed.
 One is $F_0$, for which we choose $F_0=87$MeV; the other is a cutoff $\Lambda$
 which comes from the calculation of the first term in \eqref{Snorm}.
 The details can be found in \cite{our5} and therefore we have chosen $\Lambda=$1.0$^{+0.1}_{-0.1}$GeV.
 The three-flavour numerical results are listed in the second row in Table \ref{tp4r}.
 Those LECs that depend on $\Lambda$ are expressible as
 \begin{eqnarray}
 L_{\Lambda=1\mathrm{GeV}}\bigg|^{L_{\Lambda=1.1\mathrm{GeV}}-L_{\Lambda=1\mathrm{GeV}}}_{L_{\Lambda=0.9\mathrm{GeV}}-L_{\Lambda=1\mathrm{GeV}}}.
 \end{eqnarray}
 The superscript and subscript indicate how the LECs are sensitive to $\Lambda$.
 For two flavours, we give the usual $\bar{l}_i$, $i=1,2,3,4,5,6$,
 \begin{eqnarray}
 l_i=\frac{1}{32\pi^2}\gamma_i(\bar{l}_i\!+\!\ln\frac{M^2_\pi}{\mu^2}),
 \end{eqnarray}
 where the $\gamma_i$ are given in Ref.\cite{GS1}. These results are also listed in Table \ref{tp4r}.

 \begin{table*}[!h]
 \extrarowheight 5pt
 \caption{The $p^4$ order LECs. $\Lambda_{\text{QCD}}$ is in unit of GeV, and $L_1,\ldots,L_{10},l_7$ are in units $10^{-3}$.}\label{tp4r}
 \resizebox{\textwidth}{!}{\begin{tabular}{cccccccccccc}
 \hline\hline $N_f=3$ & $\Lambda_\mathrm{QCD}$ & $L_1$
 & $L_2 $ & $L_3$ & $L_4$ & $L_5$ & $L_6$ & $L_7$ & $L_8$ & $L_9$ & $L_{10}$\\
 \hline New & 453$^{-6}_{+12}$ & $0.92^{+0.03}_{-0.04}$ & $1.84^{+0.05}_{-0.08}$ & $-4.94^{-0.14}_{+0.21}$ & $0^{+0}_{-0}$ & $1.26^{+0.01}_{-0.06}$ & $0^{+0}_{-0}$ & $-0.42^{+0.04}_{-0.05}$ & $0.84^{-0.05}_{+0.04}$ & $6.53^{+0.24}_{-0.37}$ & $-5.43^{-0.29}_{+0.44}$ \notag\\
 Old \cite{our5}&453$^{-6}_{+12}$& 1.23$^{+0.03}_{-0.04}$ &
 2.46$^{+0.05}_{-0.08}$ & -6.85$^{-0.15}_{+0.21}$ & 0$^{+0}_{-0}$ &
 1.48$^{-0.01}_{-0.03}$ & 0$^{+0}_{-0}$ & -0.51$^{+0.05}_{-0.06}$ &
 1.02$^{-0.06}_{+0.06}$&8.86$^{+0.24}_{-0.37}$ &-7.40$^{-0.29}_{+0.44}$\\
 Ref.\cite{GS2}:& &$0.9\pm 0.3$&$1.7\pm 0.7$&-$4.4\pm 2.5$&$0\pm
 0.5$&$2.2\pm 0.5$&$0\pm 0.3$&-$0.4\pm 0.15$&$1.1\pm 0.3$&$7.4\pm
 0.7$&-$6.0\pm 0.7$\\
 Ref.\cite{Pich:2008xj}:& &$0.4\pm 0.3$&$1.4\pm 0.3$&-$3.5\pm 1.1$&$-0.3\pm0.5$
 &$1.4\pm 0.5$&$-0.2\pm 0.3$&-$0.4\pm 0.2$&$0.9\pm 0.3$&$6.9\pm
 0.7$&-$5.5\pm 0.7$\\
 Ref.\cite{Colangelo:2012ipa}:& &$0.5$&$1.0$&-$3.2$&
 & & & & &$6.3$&-$6.3$\\
 Ref.\cite{gfit1}& &$0.88\pm 0.09$&$0.61\pm 0.20$&-$3.04\pm 0.43$&$0.75\pm
 0.75$&$0.58\pm 0.13$&$0.29\pm 0.85$&-$0.11\pm 0.15$&$0.18\pm 0.18$&$5.93\pm
 4.3$&-$4.06\pm 0.39$\\
 \hline $N_f=2$&$\Lambda_\mathrm{QCD}$&$\bar{l}_1$
 &$\bar{l}_2$ &$\bar{l}_3$ &$\bar{l}_4$ & $\bar{l}_5$ & $\bar{l}_6$ & $l_7$ & & &\\
 \hline New & 465$^{-6}_{+12}$ & $-2.33^{-0.17}_{+0.24}$ & $6.85^{+0.09}_{-0.14}$ & $2.33^{+0.28}_{-0.36}$ & $4.24^{+0.00}_{-0.04}$ & $13.55^{+0.53}_{-0.80}$ & $15.60^{+0.44}_{-0.67}$ & $3.61^{-0.55}_{+0.80}$ & & &\notag\\
 Old \cite{our5} &465$^{-6}_{+12}$&
 -4.77$^{-0.17}_{+0.24}$ & 8.01$^{+0.09}_{-0.14}$ & 1.97$^{+0.29}_{-0.35}$
 & 4.34$^{-0.01}_{-0.02}$ & 17.35$^{+0.53}_{-0.80}$ &
 19.98$^{+0.44}_{-0.67}$ &$4.18^{-0.65}_{+0.97}$
 \footnote{There exists a mistake in \cite{our5} for $l_7$, this is a correction.} &&&\\
 Ref.\cite{GS1}:& &$-2.3\pm 3.7$&$6.0\pm 1.3$&$2.9\pm 2.4$&$4.3\pm 0.9$&$13.9\pm 1.3$&$16.5\pm 1.1$& $O(5)$\\
 \hline\hline
 \end{tabular}}
 \end{table*}

 Comparing the ``new" and ``old" results in Table \ref{tp4r},
 the new absolute values of $L_i(\bar{l}_i)$ are, as expected, smaller than the older ones, except for $\bar{l}_3$.
 For comparison, we also list the other results obtained by different methods:
 \cite{GS1,GS2} are the first results from experimental data;
 \cite{Pich:2008xj} gives the LECs from resonance chiral theory;
 \cite{Colangelo:2012ipa} gives the LECs from a class of holographic theories;
 \cite{gfit1} gives the $L_1,\ldots,L_8$ from the global fit of the $O(p^6)$ LECs,
 and $L_{9}$ and $L_{10}$ are given in \cite{pkef,C87-3}, respectively.
 These are the usual methods to obtain LECs at present.
 On the whole, most of the old results are larger than the others, but the new results are closer.
 The only one new result larger than the old one is $\bar{l}_3$, but is also much closer than the older one.
 Our new results are much closer to the experimental results,
 and most results are within the error uncertainties of the resonance results.
 Nevertheless they appear a bit far from the results from the global fit.
 One possible reason is that the global-fitted results do not maintain the large $N_C$ limit,
 but $L_4$ and $L_6$ have fits that are not very small, the effect of which propagate throughout the calculation and decrease the values of other LECs.
 So far, our calculation remains valid only in the large $N_C$ limit, and therefore they are not very closed to the global fit results.

 These observations indicate that our approximation in \eqref{alag0} is reasonable.
 Although we only selected $n=2$ in \eqref{Snorm} and $n=1$ in \eqref{fineqNc}, the tendency is clear.
 The second and the third terms in \eqref{Snorm} carry the systematic error in our original calculations.
 To date Table \ref{tp4r} gives the leading order corrections.
 Hence, we believe that when we extend the calculations to the $O(p^6)$ order, the results will be more credible.

 \begin{table*}[!h]
 \extrarowheight 4pt
 \caption{$\tilde{Y}_n$ and their relations to $Y_i$.}\label{np6}
 \begin{eqnarray}
 \begin{array}{llllllll}
 \hline\hline n & \tilde{Y}_n & \text{relations} & n & \tilde{Y}_i & \text{relations}\\
 \hline
 1 & \la  u^{\mu}u_{\mu}h^{\nu\lambda}h_{\nu\lambda} \ra & Y_{1} & 35 & \la i f_{+}^{\mu\nu}u^{\lambda}u_{\mu}u_{\lambda}u_{\nu} +i f_{+}^{\mu\nu}u_{\mu}u^{\lambda}u_{\nu}u_{\lambda} \ra & Y_{68}\\
 2 & \la  h^{\mu\nu}u^{\lambda}h_{\mu\nu}u_{\lambda} \ra & Y_{3} & 36 & \la  u^{\mu}u_{\mu}f_{+}^{\nu\lambda}f_{+\nu\lambda} \ra & Y_{71}\\
 3 & \la  h^{\mu\nu}u^{\lambda}h_{\mu\lambda}u_{\nu} + h^{\mu\nu}u_{\nu}{h_{\mu}}^{\lambda}u_{\lambda} \ra & Y_{5} & 37 & \la  f_{+}^{\mu\nu}u^{\lambda}f_{+\mu\nu}u_{\lambda} \ra\\
 4 & \la  u^{\mu}u_{\mu}u^{\nu}u_{\nu}\chi_{+} \ra & Y_{7} & 38 & \la  f_{+}^{\mu\nu}{f_{+\mu}}^{\lambda}u_{\nu}u_{\lambda} \ra & Y_{75}\\
 5 & \la  u^{\mu}u_{\mu}u^{\nu}\chi_{+}u_{\nu} \ra & Y_{11} & 39 & \la  f_{+}^{\mu\nu}{f_{+\mu}}^{\lambda}u_{\lambda}u_{\nu} \ra & Y_{76}\\
 6 & \la  \chi_{+}u^{\mu}u^{\nu}u_{\mu}u_{\nu} \ra & Y_{13} & 40 & \la  f_{+}^{\mu\nu}u^{\lambda}f_{+\mu\lambda}u_{\nu} + f_{+}^{\mu\nu}u_{\nu}{f_{+\mu}}^{\lambda}u_{\lambda} \ra & Y_{78}\\
 7 & \la  \chi_{+}h^{\mu\nu}h_{\mu\nu} \ra & Y_{17} & 41 & \la  \chi_{+}f_{+}^{\mu\nu}f_{+\mu\nu} \ra & Y_{81}\\
 8 & \la  u^{\mu}u_{\mu}\chi_{+}\chi_{+} \ra & Y_{19} & 42 & \la i f_{+}^{\mu\nu}\chi_{+}u_{\mu}u_{\nu} +i f_{+}^{\mu\nu}u_{\mu}u_{\nu}\chi_{+} \ra & Y_{83}\\
 9 & \la  \chi_{+}u^{\mu}\chi_{+}u_{\mu} \ra & Y_{23} & 43 & \la i f_{+}^{\mu\nu}u_{\mu}\chi_{+}u_{\nu} \ra & Y_{85}\\
 10 & \la  \chi_{+}\chi_{+}\chi_{+} \ra & Y_{25} & 44 & \la  f_{-}^{\mu\nu}{h_{\nu}}^{\lambda}u_{\lambda}u_{\mu} + f_{-}^{\mu\nu}u_{\mu}u^{\lambda}h_{\nu\lambda} \ra & Y_{86}\\
 11 & \la i \chi_{-}h^{\mu\nu}u_{\mu}u_{\nu} +i \chi_{-}u^{\mu}u^{\nu}h_{\mu\nu} \ra & Y_{28} & 45 & \la  f_{-}^{\mu\nu}u_{\mu}{h_{\nu}}^{\lambda}u_{\lambda} + f_{-}^{\mu\nu}u^{\lambda}h_{\nu\lambda}u_{\mu} \ra & Y_{89}\\
 12 & \la  {h^{\lambda}}_{\lambda}h^{\mu\nu}u_{\mu}u_{\nu} + {h^{\lambda}}_{\lambda}u^{\mu}u^{\nu}h_{\mu\nu} \ra & Y_{28}-\frac{2}{N_f}Y_{30} & 46 & \la  u^{\mu}u_{\mu}f_{-}^{\nu\lambda}f_{-\nu\lambda} \ra & Y_{90}\\
 13 & \la i h^{\mu\nu}u_{\mu}\chi_{-}u_{\nu} \ra & Y_{31} & 47 & \la  f_{-}^{\mu\nu}u^{\lambda}f_{-\mu\nu}u_{\lambda} \ra & Y_{92}\\
 14 & \la  h^{\mu\nu}u_{\mu}{h^{\lambda}}_{\lambda}u_{\nu} \ra & Y_{31}-\frac{1}{N_f}Y_{30} & 48 & \la  f_{-}^{\mu\nu}{f_{-\mu}}^{\lambda}u_{\nu}u_{\lambda} \ra & Y_{94}\\
 15 & \la  u^{\mu}u_{\mu}\chi_{-}\chi_{-} \ra & Y_{33} & 49 & \la  f_{-}^{\mu\nu}{f_{-\mu}}^{\lambda}u_{\lambda}u_{\nu} \ra & Y_{95}\\
 16 & \la i u^{\mu}u_{\mu}{h^{\nu}}_{\nu}\chi_{-} +i u^{\mu}u_{\mu}\chi_{-}{h^{\nu}}_{\nu} \ra & -2Y_{33}+\frac{2}{N_f}Y_{34} & 50 & \la  f_{-}^{\mu\nu}u^{\lambda}f_{-\mu\lambda}u_{\nu} + f_{-}^{\mu\nu}u_{\nu}{f_{-\mu}}^{\lambda}u_{\lambda} \ra & Y_{97}\\
 17 & \la  u^{\mu}u_{\mu}{h^{\nu}}_{\nu}{h^{\lambda}}_{\lambda} \ra & -Y_{33}+\frac{2}{N_f}Y_{34}-\frac{1}{N_f^2}Y_{36} & 51 & \la i f_{+}^{\mu\nu}{f_{-\nu}}^{\lambda}h_{\mu\lambda} -i f_{+}^{\mu\nu}{h_{\mu}}^{\lambda}f_{-\nu\lambda} \ra & Y_{100}\\
 18 & \la  u^{\mu}\chi_{-}u_{\mu}\chi_{-} \ra & Y_{37} & 52 & \la i f_{+}^{\mu\nu}{f_{-\nu}}^{\lambda}f_{-\mu\lambda} -i f_{+}^{\mu\nu}{f_{-\mu}}^{\lambda}f_{-\nu\lambda} \ra & Y_{101}\\
 19 & \la i u^{\mu}{h^{\nu}}_{\nu}u_{\mu}\chi_{-} \ra & -Y_{37}+\frac{1}{N_f}Y_{34} & 53 & \la  \chi_{+}f_{-}^{\mu\nu}f_{-\mu\nu} \ra & Y_{102}\\
 20 & \la  u^{\mu}{h^{\nu}}_{\nu}u_{\mu}{h^{\lambda}}_{\lambda} \ra & -Y_{37}+\frac{2}{N_f}Y_{34}-\frac{1}{N_f^2}Y_{36} & 54 & \la  f_{+}^{\mu\nu}f_{-\mu\nu}\chi_{-} - f_{+}^{\mu\nu}\chi_{-}f_{-\mu\nu} \ra & Y_{104}\\
 21 & \la  \chi_{-}\chi_{-}\chi_{+} \ra & Y_{39} & 55 & \la i f_{+}^{\mu\nu}f_{-\mu\nu}{h^{\lambda}}_{\lambda} -i f_{+}^{\mu\nu}{h^{\lambda}}_{\lambda}f_{-\mu\nu} \ra & -Y_{104}\\
 22 & \la i {h^{\mu}}_{\mu}\chi_{-}\chi_{+} +i {h^{\mu}}_{\mu}\chi_{+}\chi_{-} \ra & -2Y_{39}+\frac{2}{N_f}Y_{41} & 56 & \la i f_{-}^{\mu\nu}\chi_{-}u_{\mu}u_{\nu} -i f_{-}^{\mu\nu}u_{\mu}u_{\nu}\chi_{-} \ra & Y_{105}\\
 23 & \la  {h^{\mu}}_{\mu}{h^{\nu}}_{\nu}\chi_{+} \ra & -Y_{39}+\frac{2}{N_f}Y_{41}-\frac{1}{N_f^2}Y_{42} & 57 & \la  f_{-}^{\mu\nu}{h^{\lambda}}_{\lambda}u_{\mu}u_{\nu} - f_{-}^{\mu\nu}u_{\mu}u_{\nu}{h^{\lambda}}_{\lambda} \ra & Y_{105}\\
 24 & \la i \chi_{-}\chi_{+}^{\mu}u_{\mu} +i \chi_{-}u^{\mu}\chi_{+\mu} \ra & Y_{43} & 58 & \la  f_{-}^{\mu\nu}\chi_{+\mu}u_{\nu} + f_{-}^{\mu\nu}u_{\nu}\chi_{+\mu} \ra & Y_{107}\\
 25 & \la  {h^{\nu}}_{\nu}\chi_{+}^{\mu}u_{\mu} + {h^{\nu}}_{\nu}u^{\mu}\chi_{+\mu} \ra & Y_{43}-\frac{2}{N_f}Y_{44} & 59 & \la  \nabla^{\mu}f_{-}^{\nu\lambda}\nabla_{\mu}f_{-\nu\lambda} \ra & Y_{109}\\
 26 & \la  \chi_{+}^{\mu}\chi_{+\mu} \ra & Y_{47} & 60 & \la i \nabla^{\lambda}f_{+}^{\mu\nu}h_{\mu\lambda}u_{\nu} -i \nabla^{\lambda}f_{+}^{\mu\nu}u_{\nu}h_{\mu\lambda} \ra & Y_{110}\\
 27 & \la  u^{\mu}u_{\mu}u^{\nu}u_{\nu}u^{\lambda}u_{\lambda} \ra & Y_{49} & 61 & \la i \nabla^{\mu}{f_{+\mu}}^{\nu}{f_{-\nu}}^{\lambda}u_{\lambda} -i \nabla^{\mu}{f_{+\mu}}^{\nu}u^{\lambda}f_{-\nu\lambda} \ra & Y_{111}\\
 28 & \la  u^{\mu}u_{\mu}u^{\nu}u^{\lambda}u_{\lambda}u_{\nu} \ra & Y_{52} & 62 & \la i \nabla^{\mu}{f_{+\mu}}^{\nu}{h_{\nu}}^{\lambda}u_{\lambda} -i \nabla^{\mu}{f_{+\mu}}^{\nu}u^{\lambda}h_{\nu\lambda} \ra & Y_{112}\\
 29 & \la  u^{\mu}u_{\mu}u^{\nu}u^{\lambda}u_{\nu}u_{\lambda} \ra & Y_{54} & 63 & \la i \chi_{-}^{\mu}\nabla_{\mu}{h^{\nu}}_{\nu} \ra & Z_{1}
 \footnote{$Z_1=-\frac{1}{2}Y_{19}-\frac{1}{2}Y_{23}+\frac{1}{N_f}Y_{24}-\frac{1}{2}Y_{33}+\frac{1}{N_f}Y_{34}
 -\frac{1}{2}Y_{37}-\frac{1}{2}Y_{39}
 +\frac{1}{N_f}Y_{41}-\frac{1}{2N_f^2}Y_{42}+\frac{1}{2}Y_{43}-\frac{1}{N_f}Y_{44}+\frac{1}{N_f}Y_{46}-Y_{47}+4Y_{113}$}\\
 30 & \la  u^{\mu}u^{\nu}u^{\lambda}u_{\mu}u_{\nu}u_{\lambda} \ra & Y_{58} & 64 & \la  \nabla^{\mu}{h^{\nu}}_{\nu}\nabla_{\mu}{h^{\lambda}}_{\lambda} \ra & Z_{2}
 \footnote{$Z_2=-\frac{1}{2}Y_{19}-\frac{1}{2}Y_{23}+\frac{1}{N_f}Y_{24}-Y_{33}+\frac{2}{N_f}Y_{34}-Y_{37}-Y_{39}
 +\frac{2}{N_f}Y_{41}-\frac{1}{N_f^2}Y_{42}+Y_{43}-\frac{2}{N_f}Y_{44}+\frac{1}{N_f}Y_{46}-Y_{47}+4Y_{113}$}\\
 31 & \la  u^{\mu}u^{\nu}u^{\lambda}u_{\mu}u_{\lambda}u_{\nu} \ra & Y_{60} & 65 & \la  f_{+}^{\mu\nu}u_{\mu}\chi_{-\nu} + f_{+}^{\mu\nu}\chi_{-\mu}u_{\nu} \ra & Z_{3}
 \footnote{$Z_3=-Y_{66}-Y_{67}+Y_{68}+\frac{1}{2}Y_{71}-\frac{1}{2}Y_{73}+Y_{75}-2Y_{76}+\frac{1}{2}Y_{78}
 -\frac{1}{2}Y_{83}-Y_{85}
 +\frac{1}{2}Y_{90}-\frac{1}{2}Y_{92}
 -Y_{94}+\frac{1}{2}Y_{97}-\frac{1}{2}Y_{100}+\frac{1}{2}Y_{101}-\frac{1}{4}Y_{104}+Y_{110}+Y_{112}$}\\
 32 & \la i f_{+}^{\mu\nu}u^{\lambda}u_{\lambda}u_{\mu}u_{\nu} +i f_{+}^{\mu\nu}u_{\mu}u_{\nu}u^{\lambda}u_{\lambda} \ra & Y_{64} & 66 & \la  \chi_{-}^{\mu}\chi_{-\mu} \ra & (2.15)\text{ in \cite{p62}}\\
 33 & \la i f_{+}^{\mu\nu}u^{\lambda}u_{\mu}u_{\nu}u_{\lambda} \ra & Y_{66} & 67 & \la i f_{+}^{\mu\nu}{f_{+\nu}}^{\lambda}f_{+\mu\lambda} \ra & (2.15)\text{ in \cite{p62}}\\
 34 & \la i f_{+}^{\mu\nu}u_{\mu}u^{\lambda}u_{\lambda}u_{\nu} \ra & Y_{67} & 68 & \la  \nabla^{\mu}f_{+}^{\nu\lambda}\nabla_{\mu}f_{+\nu\lambda} \ra & (2.15)\text{ in \cite{p62}}\\
 \hline\hline
 \end{array}\notag
 \end{eqnarray}
 \end{table*}

 \subsection{Order $O(p^6)$}

 Because our method only applies in the large $N_C$ limit, for simplicity, in the $O(p^6)$ order,
 we only need to calculate the large $N_C$ limit terms.
 Without the equations of motion, in large $N_C$ limit, the CL is
 \begin{eqnarray}
 \mathscr{L}_6=\sum_{n=1}^{68}\tilde{K}_n\tilde{Y}_n.
 \end{eqnarray}
 These $\tilde{Y}_n$ and their relationship to $Y_i$ defined in \cite{p62} are listed in Table \ref{np6},
 $\tilde{K}_n$ are some coefficients.

 As for the $p^2$ and $p^4$ orders, expanding \eqref{dels} to the $p^6$ order,
 with the method in Sec. \ref{cat}, treating $\tilde{Y}_n$ as $\la O_{\Omega,k}\ra$, $Y_n$ as $\la O_{l}\ra$ in \eqref{dlecs},
 we can obtain the $\Delta\tilde{K}_i$ coefficients. We have listed the values in \eqref{dkt} in Appendix \ref{tki}.
 The final $O(p^6)$ LECs are listed in Table \ref{cp6}.
 \begin{eqnarray}
 C_{\Lambda=1\mathrm{GeV}}\bigg|^{C_{\Lambda=1.1\mathrm{GeV}}-C_{\Lambda=1\mathrm{GeV}}}_{C_{\Lambda=0.9\mathrm{GeV}}-C_{\Lambda=1\mathrm{GeV}}},\hspace{1cm}
 c_{\Lambda=1\mathrm{GeV}}\bigg|^{c_{\Lambda=1.1\mathrm{GeV}}-c_{\Lambda=1\mathrm{GeV}}}_{c_{\Lambda=0.9\mathrm{GeV}}-c_{\Lambda=1\mathrm{GeV}}}\;.\notag
 \end{eqnarray}
 Because of the new relation given in Ref. \cite{p6p}, we remove $c_{37}$ as previously.
 Unlike the $O(p^4)$ order, some absolute values of the new LECs are smaller than the old ones, such as $C_1$ and $C_{4}$;
 some are almost unchanged, such as $C_3$ and $C_{12}$; some are larger than the old ones, such as $C_{52}$ and $C_{65}$;
 and some even change signs, such as $C_{22}$ and $C_{69}$.
 These arise because of the choice of the independent terms in \cite{p62} and the complex relations in \eqref{rkk}.

 The calculations are too highly complicated. To avoid possible mistakes,
 the expansion in \eqref{dels} and most of the other calculations are done by computer.
 To check the correctness of our results, we examined them in various ways.
 First, some terms in Table \ref{np6} have two parts,
 which we calculated separately. $C$, $P$, and Hermitian invariance constrains the two parts of the coefficient
 as being equal or with a sign difference.
 Our analytical results for the separate parts must give the same coefficients. Second,
 if we switch off the quark self-energy, all the LECs, except the contact terms', must vanish\cite{our5}.
 This places a strong restriction on our results.
 Third, because of the strict constraint conditions in \eqref{rkk},
 we have $109-68=41$ constraint conditions, with 109 being the number of $O_{\Omega}$ in \eqref{dlecs}.
 They also impose strong restrictions on our results.
 With all the above assessments, we are confident of the reliability of our numerical results for $O(p^6)$ LECs.

 Our choice $F_0=87$ MeV is the leading order value of $F_\pi$.
 The relation between $F_0$ and $F_\pi$ is given in Ref. \cite{piond} to the $O(p^6)$ order.
 With our results listed in Tables \ref{tp4r} and \ref{cp6}, the numerical results to the $O(p^6)$ order yield $F_\pi=92.76^{+0.01}_{-0.06}$ MeV.
 Comparing with the previous result $F_\pi=92.97^{+0.00}_{-0.04}$,
 the new result is a bit closer to that in PDG2014 \cite{pdg2014} $F_\pi=92.21$ MeV.
 From this point of view, the additional terms in \eqref{alag} improve the results slightly.

 {\extrarowheight 5pt
 \begin{longtable}{crrcrrcrrcrrcrrcrr}
 \caption{\label{cp6}The $p^6$ order LECs. They are in units of $10^{-3}$ GeV$^{-2}$.
 The value $\equiv0$ means that the LECs vanish at the lagre $N_C$ limit.}\\

 \hline\hline $i$ & new $C_i$~ & old $C_i$~ & $j$ & new $c_j$~ & old $c_j$~ &
        $i$ & new $C_i$~ & old $C_i$~ & $j$ & new $c_j$~ & old $c_j$~~~\\
 \hline\endfirsthead

 \hline\hline $i$ & new $C_i$~ & old $C_i$~ & $j$ & new $c_j$~ & old $c_j$~ &
        $i$ & new $C_i$~ & old $C_i$~ & $j$ & new $c_j$~ & old $c_j$~~~\\
 \hline\endhead

 \hline\hline 
 \endfoot

 \hline\endlastfoot

 \hline 1 & $2.98^{+0.07}_{-0.13}$ & $3.79^{+0.10}_{-0.17}$ & 1 & $3.04^{+0.07}_{-0.12}$ & $3.58^{+0.09}_{-0.15}$ & 46 & $-1.67^{-0.05}_{+0.09}$ & $-0.60^{-0.02}_{+0.04}$ & 26 & $-4.82^{-0.15}_{+0.24}$ & $-1.14^{-0.05}_{+0.07}$ \\
 2 & $\equiv 0~~~~~ $ & $\equiv 0~~~~~ $ & & &  & 47 & $3.10^{+0.09}_{-0.15}$ & $0.08^{+0.01}_{-0.00}$ & & &  \\
 3 & $-0.05^{+0.01}_{-0.01}$ & $-0.05^{+0.01}_{-0.01}$ & 2 & $-0.09^{+0.01}_{-0.01}$ & $-0.03^{+0.01}_{-0.01}$ & 48 & $4.74^{+0.10}_{-0.17}$ & $3.41^{+0.06}_{-0.10}$ & & &  \\
 4 & $2.13^{+0.06}_{-0.10}$ & $3.10^{+0.09}_{-0.15}$ & 3 & $2.15^{+0.06}_{-0.10}$ & $2.89^{+0.08}_{-0.13}$ & 49 & $\equiv 0~~~~~ $ & $\equiv 0~~~~~ $ & & &  \\
 5 & $-1.28^{+0.10}_{-0.13}$ & $-1.01^{+0.08}_{-0.11}$ & 4 & $0.82^{-0.04}_{+0.03}$ & $1.21^{-0.07}_{+0.06}$ & 50 & $10.54^{+0.89}_{-1.29}$ & $8.71^{+0.78}_{-1.12}$ & 27 & $16.70^{+1.61}_{-2.30}$ & $13.57^{+1.41}_{-2.00}$ \\
 6 & $\equiv 0~~~~~ $ & $\equiv 0~~~~~ $ & & &  & 51 & $-9.83^{+0.28}_{-0.24}$ & $-11.49^{+0.18}_{-0.09}$ & 28 & $5.51^{+1.22}_{-1.62}$ & $0.93^{+0.98}_{-1.25}$ \\
 7 & $\equiv 0~~~~~ $ & $\equiv 0~~~~~ $ & & &  & 52 & $-6.39^{-0.77}_{+1.07}$ & $-5.04^{-0.67}_{+0.93}$ & & &  \\
 8 & $2.10^{-0.15}_{+0.16}$ & $2.31^{-0.16}_{+0.18}$ & & &  & 53 & $-5.36^{-0.71}_{+1.05}$ & $-11.98^{-0.87}_{+1.33}$ & 29 & $-4.77^{-0.66}_{+0.98}$ & $-11.01^{-0.81}_{+1.23}$ \\
 9 & $\equiv 0~~~~~ $ & $\equiv 0~~~~~ $ & & &  & 54 & $\equiv 0~~~~~ $ & $\equiv 0~~~~~ $ & & &  \\
 10 & $-0.65^{+0.05}_{-0.06}$ & $-1.05^{+0.08}_{-0.09}$ & 5 & $-0.68^{+0.05}_{-0.06}$ & $-0.98^{+0.07}_{-0.09}$ & 55 & $10.45^{+0.80}_{-1.22}$ & $16.79^{+0.96}_{-1.49}$ & 30 & $9.86^{+0.75}_{-1.14}$ & $15.72^{+0.89}_{-1.38}$ \\
 11 & $\equiv 0~~~~~ $ & $\equiv 0~~~~~ $ & & &  & 56 & $4.45^{+0.03}_{-0.18}$ & $19.34^{+0.52}_{-0.98}$ & 31 & $3.27^{-0.04}_{-0.07}$ & $17.57^{+0.42}_{-0.82}$ \\
 12 & $-0.34^{+0.01}_{-0.01}$ & $-0.34^{+0.02}_{-0.01}$ & 6 & $-0.35^{+0.02}_{-0.01}$ & $-0.33^{+0.01}_{-0.01}$ & 57 & $4.72^{+1.36}_{-1.85}$ & $7.92^{+1.34}_{-1.85}$ & 32 & $4.24^{+1.32}_{-1.78}$ & $7.18^{+1.28}_{-1.76}$ \\
 13 & $\equiv 0~~~~~ $ & $\equiv 0~~~~~ $ & & &  & 58 & $\equiv 0~~~~~ $ & $\equiv 0~~~~~ $ & & &  \\
 14 & $-0.87^{+0.14}_{-0.21}$ & $-0.83^{+0.12}_{-0.19}$ & 7 & $-1.83^{+0.25}_{-0.35}$ & $-1.72^{+0.25}_{-0.35}$ & 59 & $-14.59^{-1.01}_{+1.55}$ & $-22.49^{-1.21}_{+1.89}$ & 33 & $-13.69^{-0.94}_{+1.44}$ & $-21.19^{-1.12}_{+1.76}$ \\
 15 & $\equiv 0~~~~~ $ & $\equiv 0~~~~~ $ & 8 & $0.91^{-0.11}_{+0.13}$ & $0.86^{-0.12}_{+0.15}$ & 60 & $\equiv 0~~~~~ $ & $\equiv 0~~~~~ $ & & &  \\
 16 & $\equiv 0~~~~~ $ & $\equiv 0~~~~~ $ & & &  & 61 & $2.42^{-0.19}_{+0.22}$ & $2.88^{-0.22}_{+0.26}$ & 34 & $2.40^{-0.19}_{+0.22}$ & $2.84^{-0.22}_{+0.26}$ \\
 17 & $0.17^{+0.01}_{-0.04}$ & $0.01^{-0.01}_{-0.01}$ & 9 & $-0.74^{+0.13}_{-0.18}$ & $-0.84^{+0.12}_{-0.17}$ & 62 & $\equiv 0~~~~~ $ & $\equiv 0~~~~~ $ & & &  \\
 18 & $-0.60^{+0.07}_{-0.09}$ & $-0.56^{+0.09}_{-0.11}$ & & &  & 63 & $2.48^{-0.21}_{+0.25}$ & $2.99^{-0.24}_{+0.30}$ & & &  \\
 19 & $-0.27^{+0.09}_{-0.13}$ & $-0.48^{+0.09}_{-0.13}$ & 10 & $-0.22^{+0.07}_{-0.11}$ & $-0.37^{+0.07}_{-0.10}$ & 64 & $\equiv 0~~~~~ $ & $\equiv 0~~~~~ $ & & &  \\
 20 & $0.17^{-0.02}_{+0.03}$ & $0.18^{-0.03}_{+0.04}$ & 11 & $\equiv 0~~~~~ $ & $\equiv 0~~~~~ $ & 65 & $-2.82^{+0.18}_{-0.20}$ & $-2.43^{+0.15}_{-0.16}$ & 35 & $2.16^{-0.23}_{+0.31}$ & $3.39^{-0.32}_{+0.41}$ \\
 21 & $-0.06^{+0.01}_{-0.01}$ & $-0.06^{+0.01}_{-0.01}$ & & &  & 66 & $0.80^{+0.04}_{-0.07}$ & $1.71^{+0.07}_{-0.12}$ & 36 & $0.80^{+0.04}_{-0.07}$ & $1.57^{+0.06}_{-0.10}$ \\
 22 & $-0.35^{+0.20}_{-0.26}$ & $0.27^{+0.19}_{-0.25}$ & 12 & $-0.41^{+0.20}_{-0.26}$ & $0.15^{+0.18}_{-0.24}$ & 67 & $\equiv 0~~~~~ $ & $\equiv 0~~~~~ $ &  &  &  \\
 23 & $\equiv 0~~~~~ $ & $\equiv 0~~~~~ $ & & &  & 68 & $\equiv 0~~~~~ $ & $\equiv 0~~~~~ $ & & &  \\
 24 & $0.87^{+0.02}_{-0.04}$ & $1.62^{+0.04}_{-0.07}$ & & &  & 69 & $0.52^{+0.00}_{-0.01}$ & $-0.86^{-0.04}_{+0.06}$ & 38 & $0.60^{+0.00}_{-0.01}$ & $-0.68^{-0.03}_{+0.05}$ \\
 25 & $-3.03^{-0.41}_{+0.59}$ & $-5.98^{-0.49}_{+0.72}$ & 13 & $-3.02^{-0.39}_{+0.56}$ & $-5.39^{-0.45}_{+0.66}$ & 70 & $1.66^{-0.11}_{+0.11}$ & $1.73^{-0.08}_{+0.07}$ & 39 & $1.53^{-0.12}_{+0.12}$ & $1.81^{-0.08}_{+0.07}$ \\
 26 & $2.71^{+0.35}_{-0.54}$ & $3.35^{+0.29}_{-0.47}$ & 14 & $3.39^{+0.36}_{-0.56}$ & $4.17^{+0.30}_{-0.49}$ & 71 & $\equiv 0~~~~~ $ & $\equiv 0~~~~~ $ & & &  \\
 27 & $-1.35^{+0.13}_{-0.15}$ & $-1.54^{+0.15}_{-0.18}$ & 15 & $-2.39^{+0.19}_{-0.23}$ & $-2.71^{+0.21}_{-0.25}$ & 72 & $-1.80^{+0.12}_{-0.11}$ & $-3.30^{+0.05}_{-0.00}$ & 40 & $-1.64^{+0.13}_{-0.13}$ & $-3.17^{+0.05}_{-0.02}$ \\
 28 & $0.18^{+0.00}_{-0.01}$ & $0.30^{+0.01}_{-0.01}$ & & &  & 73 & $0.15^{+0.48}_{-0.62}$ & $0.50^{+0.43}_{-0.56}$ & 41 & $0.07^{+0.47}_{-0.61}$ & $0.30^{+0.42}_{-0.54}$ \\
 29 & $-0.99^{-0.21}_{+0.24}$ & $-3.08^{-0.26}_{+0.32}$ & 16 & $-0.60^{-0.19}_{+0.21}$ & $-2.22^{-0.22}_{+0.27}$ & 74 & $-3.34^{-0.11}_{+0.19}$ & $-5.07^{-0.16}_{+0.27}$ & 42 & $-3.26^{-0.10}_{+0.17}$ & $-4.74^{-0.14}_{+0.24}$ \\
 30 & $0.37^{+0.01}_{-0.02}$ & $0.60^{+0.02}_{-0.03}$ & & &  & 75 & $\equiv 0~~~~~ $ & $\equiv 0~~~~~ $ & & &  \\
 31 & $-0.46^{+0.07}_{-0.13}$ & $-0.63^{+0.05}_{-0.09}$ & 17 & $-0.92^{+0.15}_{-0.23}$ & $-1.10^{+0.12}_{-0.19}$ & 76 & $-1.15^{-0.26}_{+0.34}$ & $-1.44^{-0.23}_{+0.31}$ & 43 & $-1.11^{-0.25}_{+0.33}$ & $-1.29^{-0.23}_{+0.30}$ \\
 32 & $0.17^{-0.02}_{+0.03}$ & $0.18^{-0.03}_{+0.04}$ & 18 & $0.42^{-0.06}_{+0.08}$ & $0.43^{-0.07}_{+0.08}$ & 77 & $\equiv 0~~~~~ $ & $\equiv 0~~~~~ $ & & &  \\
 33 & $-0.05^{-0.02}_{+0.05}$ & $0.09^{-0.00}_{+0.03}$ & 19 & $0.29^{-0.07}_{+0.13}$ & $0.41^{-0.06}_{+0.10}$ & 78 & $8.82^{+0.80}_{-1.21}$ & $17.51^{+1.02}_{-1.59}$ & 44 & $8.19^{+0.74}_{-1.12}$ & $16.16^{+0.94}_{-1.45}$ \\
 34 & $0.66^{-0.18}_{+0.29}$ & $1.59^{-0.10}_{+0.17}$ & 20 & $0.74^{-0.18}_{+0.28}$ & $1.56^{-0.10}_{+0.17}$ & 79 & $5.86^{-0.14}_{+0.12}$ & $-0.56^{-0.30}_{+0.40}$ & 45 & $6.09^{-0.13}_{+0.11}$ & $0.26^{-0.26}_{+0.34}$ \\
 35 & $0.10^{-0.09}_{+0.12}$ & $0.17^{-0.12}_{+0.17}$ & 21 & $0.22^{-0.14}_{+0.19}$ & $0.29^{-0.18}_{+0.24}$ & 80 & $1.01^{-0.05}_{+0.04}$ & $0.87^{-0.04}_{+0.03}$ & 46 & $1.09^{-0.05}_{+0.05}$ & $0.85^{-0.04}_{+0.02}$ \\
 36 & $\equiv 0~~~~~ $ & $\equiv 0~~~~~ $ & & &  & 81 & $\equiv 0~~~~~ $ & $\equiv 0~~~~~ $ & & &  \\
 37 & $-0.60^{+0.07}_{-0.09}$ & $-0.56^{+0.09}_{-0.11}$ & & &  & 82 & $-4.58^{-0.24}_{+0.38}$ & $-7.13^{-0.32}_{+0.51}$ & 47 & $-4.26^{-0.22}_{+0.35}$ & $-6.73^{-0.29}_{+0.47}$ \\
 38 & $0.47^{-0.04}_{+0.02}$ & $0.41^{-0.08}_{+0.07}$ & 22 & $-1.27^{+0.19}_{-0.26}$ & $-1.32^{+0.18}_{-0.25}$ & 83 & $-1.74^{+0.17}_{-0.22}$ & $0.07^{+0.20}_{-0.27}$ & 48 & $-1.71^{+0.17}_{-0.21}$ & $-0.22^{+0.18}_{-0.25}$ \\
 39 & $\equiv 0~~~~~ $ & $\equiv 0~~~~~ $ & 23 & $0.91^{-0.11}_{+0.13}$ & $0.86^{-0.12}_{+0.15}$ & 84 & $\equiv 0~~~~~ $ & $\equiv 0~~~~~ $ & & &  \\
 40 & $-4.98^{-0.14}_{+0.25}$ & $-6.35^{-0.18}_{+0.32}$ & 24 & $0.00^{-0.01}_{+0.03}$ & $-4.84^{-0.14}_{+0.25}$ & 85 & $-0.96^{+0.04}_{-0.03}$ & $-0.82^{+0.03}_{-0.02}$ & 49 & $-1.07^{+0.05}_{-0.04}$ & $-0.78^{+0.03}_{-0.01}$ \\
 41 & $\equiv 0~~~~~ $ & $\equiv 0~~~~~ $ & & &  & 86 & $\equiv 0~~~~~ $ & $\equiv 0~~~~~ $ & & &  \\
 42 & $1.88^{+0.03}_{-0.06}$ & $0.60^{-0.00}_{+0.00}$ & & &  & 87 & $4.79^{+0.29}_{-0.46}$ & $7.57^{+0.37}_{-0.60}$ & 50 & $4.35^{+0.26}_{-0.42}$ & $7.18^{+0.34}_{-0.55}$ \\
 43 & $\equiv 0~~~~~ $ & $\equiv 0~~~~~ $ & & &  & 88 & $-1.69^{-0.68}_{+0.93}$ & $-5.47^{-0.73}_{+1.03}$ & 51 & $-1.36^{-0.65}_{+0.89}$ & $-4.85^{-0.69}_{+0.97}$ \\
 44 & $1.73^{+0.07}_{-0.14}$ & $6.32^{+0.20}_{-0.36}$ & 25 & $4.88^{+0.16}_{-0.27}$ & $6.03^{+0.19}_{-0.33}$ & 89 & $17.27^{+1.11}_{-1.77}$ & $34.74^{+1.61}_{-2.62}$ & 52 & $15.84^{+1.00}_{-1.60}$ & $32.19^{+1.46}_{-2.37}$ \\
 45 & $\equiv 0~~~~~ $ & $\equiv 0~~~~~ $ & & &  & 90 & $2.32^{-0.44}_{+0.55}$ & $2.44^{-0.38}_{+0.46}$ & 53 & $2.28^{-0.44}_{+0.55}$ & $2.51^{-0.37}_{+0.46}$ \\
 \hline\hline
 \end{longtable}
 }

 \subsection{Anomaly}
 Following the same procedures as for the normal parts, the anomalous LECs can also be revised.
 The $O(p^4)$ CL are Wess-Zumino terms, that had been obtained from the first term in \eqref{Snorm} \cite{oura1,oura2}.
 The additional terms in \eqref{alag} must vanish to the $O(p^4)$ order.,
 thereby imposing another requirement.
 We checked our calculations and verified this requirement.
 From another point of view, we checked that the terms with $n=2$ in \eqref{Snorm} and $n=1$ in \eqref{fineqNc} were suitable.

 To the $O(p^6)$ order, without the equations of motion, in the large $N_C$ limit,
 the $n$-flavor CL is
 \begin{eqnarray}
 \mathscr{L}^W_6=\sum_{n=1}^{23}\tilde{K}^W_n\tilde{O}^W_n.
 \end{eqnarray}
 These $\tilde{O}^W_n$ and their relations with $O^W_I$ in Ref. \cite{p6a2} are listed in Table VI in Ref. \cite{oura2}.
 The analytical results for $\Delta\tilde{K}^W_n$ are listed in \eqref{dkwt} in Appendix \ref{tobi}.
 $C^W_3$ $C^W_{18}$ and $c^W_{12}$ vanish in the large $N_C$ limit.
 The nonzero numerical results are listed in the second column in Table \ref{ap63} for three flavors and
 in the second row in Table \ref{ap62} for two flavours.
 \begin{eqnarray}
 C^W_{\Lambda=1\mathrm{GeV}}\bigg|^{C^W_{\Lambda=1.1\mathrm{GeV}}-C^W_{\Lambda=1\mathrm{GeV}}}_{C^W_{\Lambda=0.9\mathrm{GeV}}-C^W_{\Lambda=1\mathrm{GeV}}},\hspace{1cm}
 c^W_{\Lambda=1\mathrm{GeV}}\bigg|^{c^W_{\Lambda=1.1\mathrm{GeV}}-c^W_{\Lambda=1\mathrm{GeV}}}_{c^W_{\Lambda=0.9\mathrm{GeV}}-c^W_{\Lambda=1\mathrm{GeV}}}\;.\notag
 \end{eqnarray}

 \begin{table*}[!h]
 \caption{The nonzero values of the $p^6$ order anomalous LECs $C^W_i$ in three flavors.
 They are in units of $10^{-3}$ GeV$^{-2}$.
 The forth column to the eighth column contain the results given in \cite{Strandberg:2003zf}: (I)-ChPT, (II)-VMD, (III)-ChPT
 (extrapolation), (IV)-CQM, (V)-CQM (extrapolation).}\label{ap63}
 \resizebox{\textwidth}{!}{
 \begin{tabular}{crrcccccrcccc}
 \hline\hline $n$& new~~~~ & old\cite{oura2}~~ & (I)\cite{Strandberg:2003zf} & (II)\cite{Strandberg:2003zf}
  & (III)\cite{Strandberg:2003zf} & (IV)\cite{Strandberg:2003zf} & (V)\cite{Strandberg:2003zf}
   & \cite{Colangelo:2012ipa} & \cite{PiDecay,olecs1,olecs2,olecs3,olecs4,hola1}\\
 \hline 1 & $2.90^{+0.49}_{-0.69}$ & $4.97^{+0.55}_{-0.79}$& & & & & & &\\
 2 & $-1.79^{+0.09}_{-0.11}$ & $-1.43^{+0.10}_{-0.12}$& $-0.32\pm10.4$  && $0.78\pm 12.7$ & $4.96\pm9.70$ & $-0.074\pm13.3$& &\\
 4 & $-1.89^{+0.19}_{-0.24}$ & $-0.96^{+0.22}_{-0.29}$& $0.28\pm9.19$   & &$0.67\pm 10.9$ & $6.32\pm6.09$ & $-0.55\pm9.05$& &\\
 5 & $1.56^{+0.29}_{-0.41}$ & $3.26^{+0.34}_{-0.49}$& $28.5\pm28.83$  & &$9.38\pm 152.2$ & $33.05\pm28.66$& $34.51\pm41.13$& &\\
 6 & $0.72^{+0.02}_{-0.04}$ & $0.91^{+0.03}_{-0.04}$& & & & & & &\\
 7 & $2.02^{-0.23}_{+0.29}$ & $1.68^{-0.24}_{+0.31}$& $0.013\pm1.17$  & & & $0.51\pm0.06$ & & &$0.1\pm1.2$\\
 & & & $20.3\pm18.7$   & & & && &$1.0^a$\\
 & & & & & & && &$0.35\pm0.07$\\
 8 & $0.52^{+0.01}_{-0.03}$ & $0.41^{+0.01}_{-0.02}$& $0.76\pm0.18$   & & & & & &$0.58\pm0.20$\\
 & & &   & & & & & &$5.0^a$\\
 9 & $1.21^{-0.03}_{+0.02}$ & $1.15^{-0.03}_{+0.03}$&  & & & & & &\\
 10 & $-0.14^{-0.00}_{+0.01}$ & $-0.18^{-0.01}_{+0.01}$&  & & & & & &\\
 11 & $-1.39^{+0.07}_{-0.09}$ & $-1.15^{+0.08}_{-0.10}$&$ -6.37\pm4.54$  & & & $-0.00143\pm0.03$&  &&$0.68\pm0.21$\\
 12 & $-4.05^{-0.12}_{+0.20}$ & $-5.13^{-0.15}_{+0.25}$&  & & & & & -2.1&\\
 13 & $-6.81^{-0.20}_{+0.33}$ & $-6.37^{-0.18}_{+0.31}$& $-74.09\pm55.89$& -20.00 & $-8.44\pm69.9$  & $14.15\pm 15.22$ & $-7.46\pm19.62$& -8.8&\\
 14 & $-2.48^{-0.07}_{+0.12}$ & $-2.00^{-0.06}_{+0.10}$& $29.99\pm11.14$ & -6.01  & $0.72\pm15.3$  & $10.23\pm7.56$ & $-0.58\pm9.77$& -1.3&\\
 15 & $2.35^{+0.07}_{-0.11}$ & $4.17^{+0.12}_{-0.20}$& $-25.3\pm 23.93$& 2.00   & $-3.10\pm28.6$ & $19.70\pm7.49$ & $8.89\pm9.72$& 4.4&\\
 16 & $1.79^{+0.05}_{-0.09}$ & $3.58^{+0.10}_{-0.17}$& & &  & && -0.2&\\
 17 & $0.71^{+0.02}_{-0.03}$ & $1.98^{+0.06}_{-0.10}$& & &  & && -0.1&\\
 19 & $0.56^{+0.02}_{-0.03}$ & $0.29^{+0.01}_{-0.01}$& & &  & && -7.0&\\
 20 & $-1.02^{-0.03}_{+0.05}$ & $1.82^{+0.05}_{-0.09}$& & &  & && -0.4&\\
 21 & $3.11^{+0.09}_{-0.15}$ & $2.48^{+0.07}_{-0.12}$& & &  & && 2.6&\\
 22 & $4.72^{+0.14}_{-0.23}$ & $5.01^{+0.14}_{-0.24}$& $6.52\pm0.78$   & 8.01 &  & $3.94\pm0.43$ && 7.9&$5.4\pm0.8$\\
 & & & $5.07\pm0.71$   & & & $3.94 \pm0.43$ && &6.71,6.21,4.45$^b$\\
 23 & $2.80^{+0.08}_{-0.14}$ & $2.74^{+0.08}_{-0.13}$& & &  & && 0.9&\\
 \hline\hline
 \end{tabular}
 }
 \begin{flushleft}
 $^a$ This result is just the absolute value given in \cite{olecs3}.

 $^b$ These result are in \cite{hola1} by different inputs.
 \end{flushleft}
 \end{table*}

 \begin{table*}[!h]
 \extrarowheight 4pt
 \caption{The nonzero values of the $p^6$ order anomalous LECs $c^W_i$ in two flavors.
 They are in units of $10^{-3}$ GeV$^{-2}$.}\label{ap62}
 \resizebox{\textwidth}{!}{\begin{tabular}{ccccccccccccc}
 \hline\hline &$c_1^W$ & $c_2^W$ & $c_3^W$& $c_4^W$& $c_5^W$& $c_6^W$& $c_7^W$& c$_8^W$& $c_9^W$& $c_{10}^W$& $c_{11}^W$& $c_{13}^W$\\
 \hline\text{new}&$-1.81^{+0.09}_{-0.11}$ &  $-1.61^{+0.08}_{-0.09}$ &  $3.61^{-0.19}_{+0.22}$ &  $0.80^{-0.04}_{+0.04}$ &  $-1.40^{+0.07}_{-0.09}$ &  $0.52^{+0.26}_{-0.35}$ &  $-0.41^{-0.01}_{+0.02}$ &  $0.21^{+0.01}_{-0.01}$ &  $6.49^{+0.18}_{-0.30}$ &  $-6.15^{-0.17}_{+0.29}$ &  $4.63^{+0.13}_{-0.22}$ &    $-9.25^{-0.26}_{+0.43}$ \notag\\
 \text{old}\cite{oura2}&$-1.46^{+0.10}_{-0.12}$ &  $-1.25^{+0.09}_{-0.11}$ &  $2.96^{-0.20}_{+0.25}$ &  $0.63^{-0.04}_{+0.05}$ &  $-1.17^{+0.08}_{-0.10}$ &  $0.77^{+0.26}_{-0.36}$ &  $-0.04^{-0.00}_{+0.00}$ &  $0.02^{+0.00}_{-0.00}$ &  $8.19^{+0.23}_{-0.38}$ &  $-8.73^{-0.24}_{+0.41}$ &  $4.85^{+0.13}_{-0.23}$ &    $-9.70^{-0.27}_{+0.45}$\notag\\
 \hline\hline
 \end{tabular}}
 \end{table*}

 \section{Comparisons}\label{comp}
 In this section, we shall gather the LECs to the $O(p^6)$ order given in the literature to provide a means to assess our new results.
 It is just only an update of \cite{our5} but also includes some new results.
 Usually, these LECs are given as dimensionless parameters with the convention of $C_i^r\equiv C_iF_0^2$ or $c_i^r\equiv c_iF_0^2$ .
 The following values of the physical constants come from the central values of PDG2014 \cite{pdg2014},
 \begin{eqnarray}
 m_{\pi^\pm}=139.57018\text{MeV},m_{\pi^0}=134.9766\text{MeV},F_\pi=92.21\text{MeV},
 m_{K^\pm}=493.677\text{MeV}.\label{expd}
 \end{eqnarray}
 Some of these values are needed in certain parts of the calculations.

 \subsection{$\pi\pi$ and $\pi K$ scattering}
 In $\pi\pi$ scattering there exist six additional constants, $r^r_i,i=1,2,3,4,5,6$.
 Their relationship to LECs can be found in Eq. (5.3) in \cite{p6rn}.
 Ref. \cite{p6rn} also gives their values obtained using two theoretical methods,
 resonance-saturation (RS)\cite{Bijnens:pp} and pure dimensional analysis
 (ND) which only accounts for the order of magnitude.
 Ref. \cite{pp3,pp4} also give the $r^r_i$ parameters,
 all of which are listed in Table. \ref{pp2}.
 Our new $r^r_{4,5,6}$ only change slightly, the error bars from the references being also small.
 They are nonetheless in good agreement. However, in the references, $r^r_{1,2,3}$ have large error bars,
 and hence our new results can potentially change signs.
 \begin{table*}[!h]
 \extrarowheight 4pt
 \caption{The obtained values for the combinations of the $p^6$ order LECs
 from $\pi\pi$ scattering and our work. The coefficients in the table are in units of $10^{-4}$.}\label{pp2}
 \begin{eqnarray*}
 \begin{array}{lcccccc}
 \hline\hline  & r^r_1 & r^r_2 & r^r_3&r^r_4&r^r_5 & r^r_6\\
 \hline \mbox{RS in Ref.\cite{p6rn}} & -0.6&1.3&-1.7&-1.0&1.1&0.3\\
 \mbox{ND in Ref.\cite{p6rn}} & 80&40&20&3&6&2\\
 \mbox{set C (n=5)\cite{pp3}}&-14\pm17\pm3&22\pm16\pm4&-3\pm1\pm3&-0.22\pm0.13\pm0.05&0.9\pm0.1\pm0.5&0.25\pm0.01\pm0.05\\
 \mbox{set C (n=3)\cite{pp3}}&-20\pm17\pm3&7\pm10\pm4&-4\pm1\pm3&0.13\pm0.13\pm0.05&0.9\pm0.1\pm0.5&0.25\pm0.01\pm0.05\\
 \mbox{Ref. \cite{pp4}}&&18&0.9&-1.9&&\\
 \mbox{Old\cite{our5}}  & -9.32^{-2.62}_{+3.51} & 8.93^{+3.12}_{-4.27} & -3.06^{-0.81}_{+1.11} &
 -0.12^{+0.22}_{-0.29}&0.87^{+0.04}_{-0.06}&0.42^{+0.02}_{-0.03}\\
 \text{new} & 0.11^{-2.50}_{+3.27} & -2.84^{+2.95}_{-3.94} & 1.03^{-0.69}_{+0.92} & -0.63^{+0.21}_{-0.28} & 0.37^{+0.02}_{-0.04} & 0.28^{+0.01}_{-0.02}\\
 \hline\hline
 \end{array}
 \end{eqnarray*}
 \end{table*}

 Furthermore, Ref. \cite{pk} introduces some coefficients, such as $c^+_{20}$, $c^+_{01}$, $c^+_{10}$, in $\pi K$ Scattering.
 Table \ref{pkb} lists all these coefficients and some LECs which were obtained in Ref. \cite{pk1} from different models.
 Whereas the new results on the left-hand side of Table \ref{pkb} are smaller and approach their results,
 results on the right-hand side of of Table \ref{pkb} seem worse.
 This may be because of propagation errors from the complex relations between $c^+$ and $C_i$.
 \begin{table*}[!h]
 \extrarowheight 4pt
 \caption{The obtained values for the combinations of the $p^6$ order LECs
 from $\pi K$ scattering and our work.
 The LECs in the l.h.s. of the
 table are in units of $10^{-4}\mathrm{GeV}^{-2}$.}\label{pkb}
 \begin{eqnarray*}
 \begin{array}{lcccc|lccc}
 \hline\hline  & C_1+4C_3 & C_2 & C_4+3C_3 & C_1+4C_3+2C_2&&
 c_{20}^+\frac{m_\pi^4}{F_\pi^4}&c_{01}^+\frac{m_\pi^2}{F_\pi^4}&c_{10}^-\frac{m_\pi^3}{F_\pi^4}\\
 \hline \mbox{input~}c^+_{30},c^+_{11},c^-_{20} & 20.7\pm4.9 & -9.2\pm4.9 & 9.9\pm2.5  & 2.3\pm10.8&&&\\
 \mbox{input~}c^+_{30},c^+_{11},c^-_{01} & 28.1\pm4.9 & -7.4\pm4.9 & 21.0\pm2.5 & 13.4\pm10.8
 &\mbox{Dispersive}&\hspace*{-1cm}0.024\pm0.006&2.07\pm0.10&0.31\pm0.01\\
 \pi \pi \mbox{~amplitude}                           &            &            & 23.5\pm2.3 & 18.8\pm7.2&&&\\
 \mbox{Resonance model}                  & 7.2        & -0.5       & 10.0       & 6.2&\mbox{Resonance~model}
 &0.003&3.8&0.09\\
 \mbox{Old \cite{our5}} & 35.9^{+1.3}_{-2.1} & 0.0^{+0.0}_{-0.0} & 29.5^{+1.1}_{-1.9}
 & 35.9^{+1.3}_{-2.1}& \mbox{Old \cite{our5}}&\hspace*{-0.5cm}0.006^{-0.002}_{+0.003}&-0.159^{+0.133}_{-0.178}&0.020^{+0.037}_{-0.050}\\
 \mbox{New} & 27.8^{+1.1}_{-1.8} & 0.0^{+0.0}_{-0.0} & 19.8^{+0.9}_{-1.4}
 & 27.8^{+1.1}_{-1.8}& \mbox{New}&\hspace*{-0.5cm}0.016^{-0.002}_{+0.002}&-0.474^{+0.142}_{-0.186}&-0.082^{+0.036}_{-0.047}\\
 \hline\hline
 \end{array}
 \end{eqnarray*}
 \end{table*}

 \subsection{Form factors}
 Ref. \cite{p6rn} also estimates the expressions of the vector form factor, the scalar form factor,
 two form factors of $\pi(p)\to e\nu\gamma(q)$ with LECs,
 and \cite{Kl3,Kpi} give some LECs using measurements of the pion scalar form factor, $K_{l3}$
 and the $\pi K$ form factors.
 Ref. \cite{pp3} gives one form factor from $\pi\pi$-scattering.
 All of them are listed in Table \ref{ff}.
 Ref. \cite{kpif} extrapolates the lattice data on the scalar $K\pi$ form factor to obtain some LECs;
 the results are listed in Table \ref{kpi}.
 Two of these results, $r^r_{S2}$ and $2C_{12}^r+C_{34}^r$ , seem better, as their signs have been changed.

\begin{figure}[h]
 \begin{minipage}[b]{\textwidth}
 \includegraphics[scale=0.6]{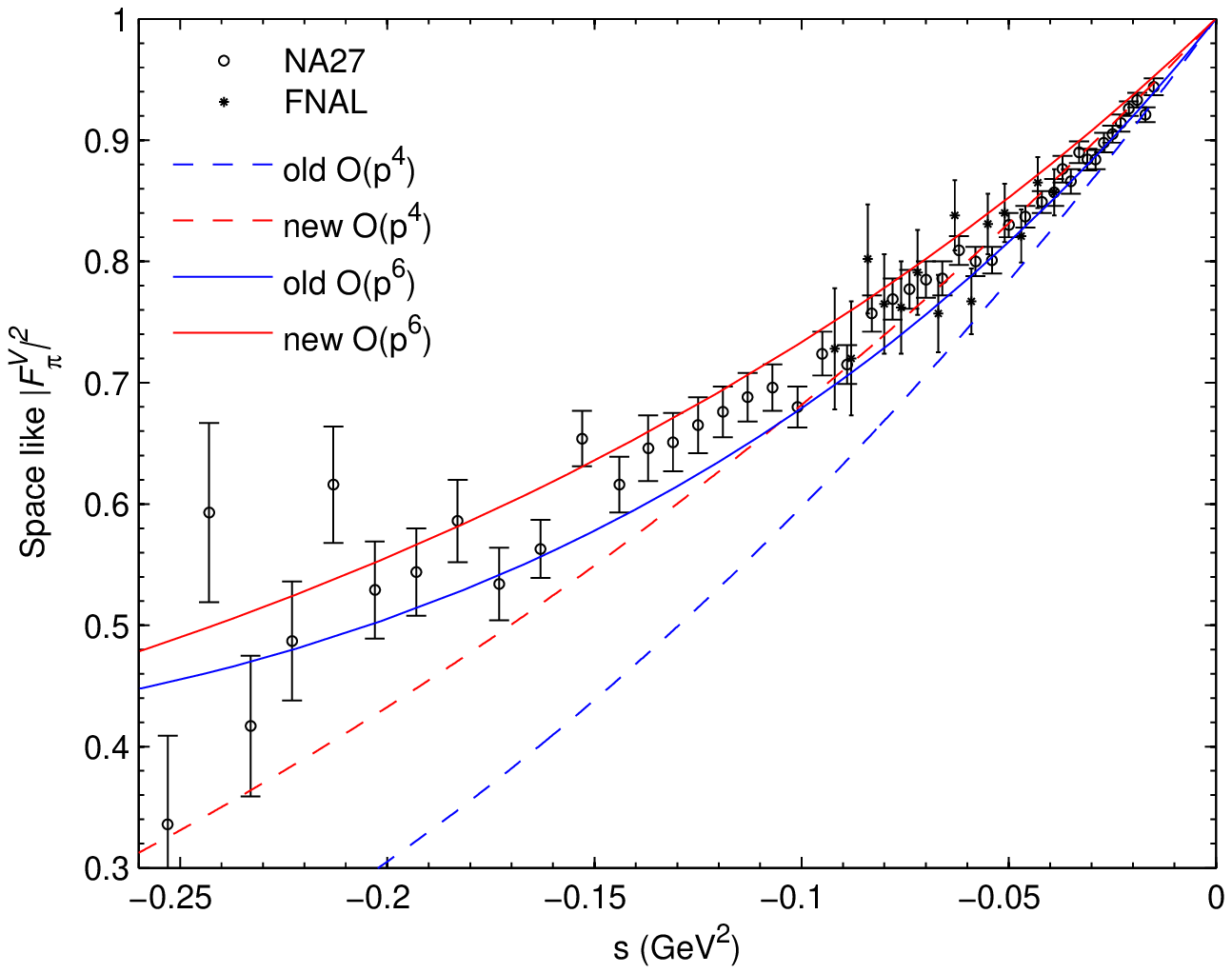}\includegraphics[scale=0.6]{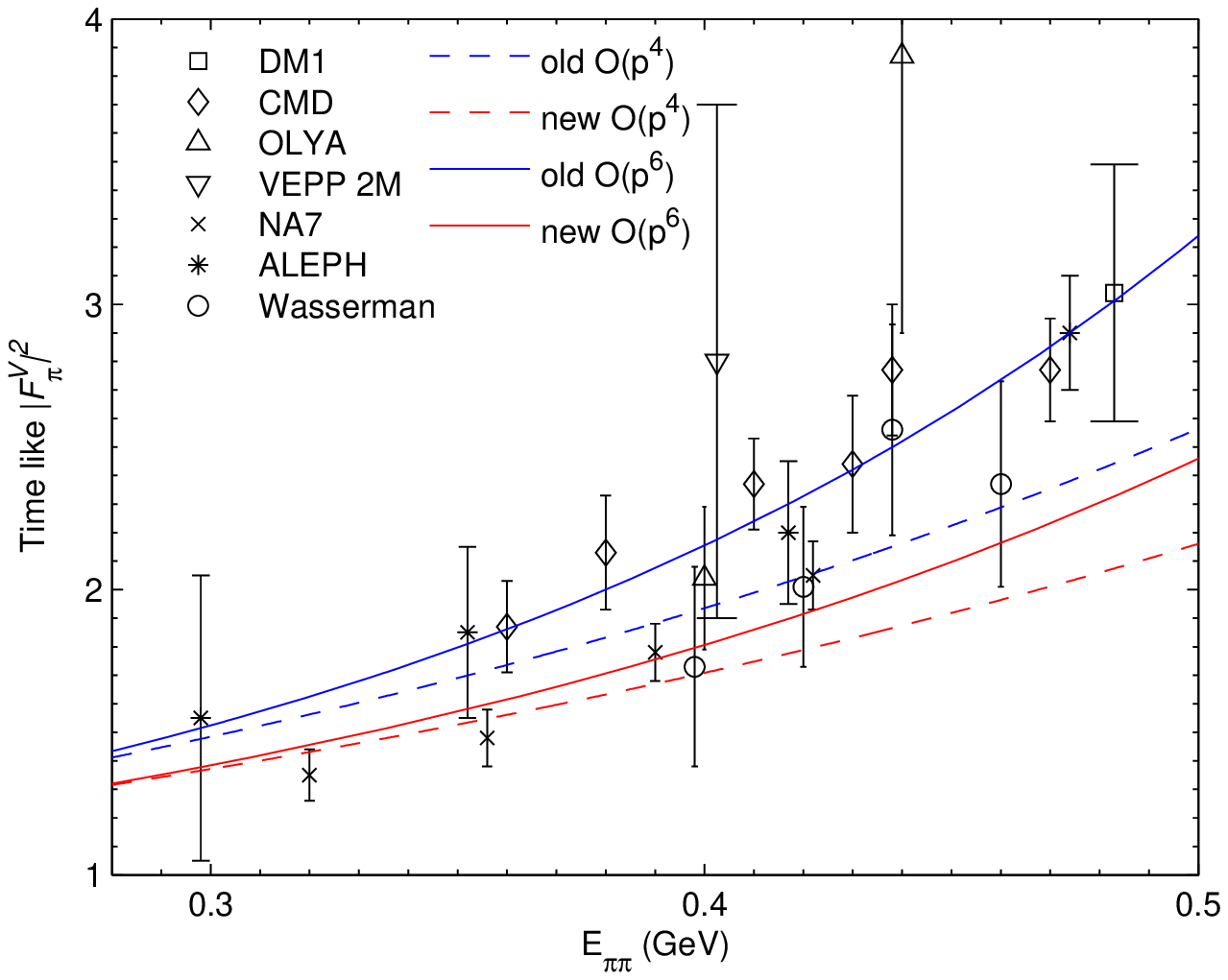}
 \end{minipage}
 \caption{The space like and time like data for the vector form factor with the old and new results.} \label{Vecf}
 \end{figure}

 Furthermore in Fig.\ref{Vecf}, we compare the experimental data \cite{FNAL,NA7,OLYA,ALEPH,DM1,TOF,NA72} for the vector form factors collected in Figure 4
 and 5 in \cite{piond} with our old and new results. In obtaining our numerical predictions,
 we have exploited the formula given by Eq.(3.16) in \cite{piond} which especially depends on $O(p^6)$
 LECs through $r_{V1}^r$ and $r_{V2}^r$ defined in \cite{p6rn}, and we
 input the formula with the old and new $O(p^4)$ and $O(p^6)$ LECs.

 From Fig.\ref{Vecf}, we see that at both $O(p^4)$ and $O(p^6)$,
 for the space-like form factors, the new line is higher than the old ones, whereas
 for the time-like form factors, the new line are lower than old ones.
 The new results are slightly worse than the old ones.
 Nevertheless, considering the experimental errors, they are all consistent with the experimental data.

 \begin{table*}[!h]
 \extrarowheight 4pt
 \caption{The obtained values for the combinations of the $p^6$ order LECs
 appear in vector and scalar form factor of pion.
 the coefficients in the table are in units of $10^{-4}$.}\label{ff}
 \begin{eqnarray*}
 &\begin{array}{cccc|ccccc|cccc}
 \hline\hline & \text{New} & \mbox{Old\cite{our5}} & \mbox{Ref.\cite{p6rn}} & & \text{New}&\mbox{Old\cite{our5}} & \mbox{Ref.\cite{p6rn}}& \mbox{Ref.\cite{pp3}}& & \text{New}&\mbox{Old\cite{our5}}
 & \mbox{Ref.\cite{p6rn}}\\
 \hline r^r_{V1} & -1.60^{+0.32}_{-0.41} & -2.13^{+0.30}_{-0.39} & -2.5 & r^r_{S2}& -0.86^{+0.00}_{+0.00}& 0.07^{+0.05}_{-0.08}&-0.3&1\pm4\pm1 &r_{A1}^r& 1.29^{+0.05}_{-0.06}&1.14^{+0.07}_{-0.09}&-0.5\\
  r^r_{V2} & 1.10^{+0.07}_{-0.10}& 2.23^{+0.10}_{-0.16} & 2.6  & r^r_{S3}& 0.21^{-0.01}_{+0.01}&0.20^{-0.01}_{+0.01}&0.6&&r_{A2}^r& -0.72^{-0.07}_{+0.10}&-0.38^{-0.06}_{+0.08}&1.1\\
 \hline\hline
 \end{array}&\notag\\
 &\begin{array}{cccc|cccc}
 \hline\hline  & \text{New}&\mbox{Old\cite{our5}}&\mbox{Ref.\cite{Kl3}}&& \text{New}& \mbox{Old\cite{our5}}&\mbox{Ref.\cite{Kpi}}\\
 \hline C_{12}^r& -0.026^{+0.001}_{-0.001}&-0.026^{+0.001}_{-0.001}&-0.1&C^r_{12}& -0.026^{+0.001}_{-0.001}&
 -0.026^{+0.001}_{-0.001}&(0.3\pm 5.4)\times 10^{-3}\\
  2C_{12}^r\!\!+C_{34}^r& -0.001^{-0.012}_{+0.020}&
 0.068^{-0.006}_{+0.010}&-0.10\pm 0.17&C_{12}^r\!\!+C_{34}^r& 0.025^{-0.013}_{+0.021}&0.094^{-0.007}_{+0.011}&(3.2\pm 1.5)\times 10^{-2}\\
 \hline\hline
 \end{array}&
 \end{eqnarray*}
 \end{table*}

 \begin{table*}[!h]
 \extrarowheight 4pt
 \caption{The LECs come from extrapolating the lattice data on the scalar $K\pi$ form factor.
 They are in units of $10^{-4}$GeV$^{-2}$.}\label{kpi}
 \begin{eqnarray*}
 \begin{array}{lcccccc}
 \hline\hline  &C_{12}&C_{34}&C_{14}&2C_{17}\\
 \hline \mbox{Ref.\cite{kpif}} & 5.74\pm0.95& 1.07\pm0.96&0.71\pm1.42&1.92\pm3.36\\
 \mbox{Old\cite{our5}}& -0.34^{+0.02}_{-0.01} & 1.59^{-0.10}_{+0.17} & -0.83^{+0.12}_{-0.19} & 0.03^{-0.02}_{-0.02}\\
 \text{New}& -0.34^{+0.01}_{-0.01} & 0.66^{-0.18}_{+0.29} & -0.87^{+0.14}_{-0.21} & 0.35^{+0.03}_{-0.08}\\
 \hline\hline
 \end{array}
 \end{eqnarray*}
 \end{table*}

 \subsection{Photon-photon collisions}
 Ref. \cite{pigamma,pigamma1} introduce some parameters by $\gamma\gamma\to\pi^0\pi^0$ and $\gamma\gamma\to\pi^+\pi^-$.
 These parameters are also all related to LECs. They are listed in Table \ref{gg}.
 These results are nearly unchanged, with $a^r_2$ and $b^r$ still having opposite signs.
 \begin{table*}[!h]
 \extrarowheight 4pt
 \caption{The obtained values for the combinations of the $p^6$ order LECs
 appear in photon-photon collisions.}\label{gg}
 \begin{eqnarray*}
 \begin{array}{cccc|cccc}
 \hline\hline & \text{New}& \mbox{Old\cite{our5}} & \mbox{Ref.\cite{pigamma}} & & \text{New}&\mbox{Old\cite{our5}} & \mbox{Ref.\cite{pigamma1}}\\
 \hline a^r_1 & -8.15^{-0.85}_{+1.17}& -5.65^{-0.91}_{+1.23} & -14\pm 5 & a^r_1& -5.46^{-0.12}_{+0.07}& -5.86^{-0.49}_{+0.58}&-3.2\\
  a^r_2 & 3.94^{+0.03}_{-0.06}& 3.79^{+0.02}_{-0.05} & 7\pm 3  & a^r_2& -1.73^{-0.09}_{+0.16}&-0.98^{-0.07}_{+0.12}&0.7\\
  b^r & 1.87^{+0.06}_{-0.10}& 1.66^{+0.05}_{-0.09} & 3\pm 1  & b^r& -0.20^{-0.02}_{+0.02}&-0.23^{-0.01}_{+0.02}&0.4\\
 \hline\hline
 \end{array}
 \end{eqnarray*}
 \end{table*}

 \subsection{Radiative pion decay}
 Ref. \cite{PiDecay} gives a group of LECs to the $O(p^6)$ order. They are listed in Table \ref{rpd}.
 Most of the new results are fitted better than the old ones, with the exception of $C^r_{88}$.
 \begin{table*}[!h]
 \extrarowheight 4pt
 \caption{The obtained values for the combinations of the $p^6$ order LECs
 from pion radiative decay and our work.
 The coefficients in the table are in units of $10^{-5}$.}\label{rpd}
 \begin{eqnarray*}
 \begin{array}{lcccccc}
 \hline\hline  &C^r_{12}&C^r_{13}&C^r_{61}&C^r_{62}&2C^r_{63}-C^r_{65}&C^r_{64}\\
 \hline \mbox{Ref.\cite{PiDecay}} & -0.6\pm 0.3&0\pm 0.2&1.0\pm 0.3& 0\pm 0.2&1.8\pm 0.7 &0\pm 0.2\\
 \hline
 \mbox{Old\cite{our5}}&-0.26^{+0.01}_{-0.01}&0.0^{+0.0}_{-0.0}&2.18^{-0.17}_{+0.20}&0.0^{+0.0}_{-0.0}&6.36^{-0.42}_{+0.56}
 &0.0^{+0.0}_{-0.0}\\
 \text{New}& -0.26^{+0.01}_{-0.01} & 0.00^{+0.00}_{+0.00} & 1.83^{-0.14}_{+0.17} & 0.00^{+0.00}_{+0.00} & 5.89^{-0.45}_{+0.53} & 0.00^{+0.00}_{+0.00}\\
 \hline\hline
 &C^r_{78}&C^r_{80}&C^r_{81}&C^r_{82}& C^r_{87}&C^r_{88}\\
 \hline  \mbox{Ref.\cite{PiDecay}}&10.0\pm 3.0&1.8\pm 0.4&
 0\pm 0.2& -3.5\pm 1.0&3.6\pm 1.0 &-3.5\pm 1.0\\
 \hline
 \mbox{Old\cite{our5}}&13.26^{+0.77}_{-1.20}&0.66^{-0.03}_{+0.02}&0.0^{+0.0}_{-0.0}&-5.39^{-0.24}_{+0.39}&5.73^{+0.28}_{-0.45}
 &-4.14^{-0.55}_{+0.78}\\
 \text{New}& 6.68^{+0.60}_{-0.91} & 0.77^{-0.04}_{+0.03} & 0.00^{+0.00}_{+0.00} & -3.47^{-0.18}_{+0.29} & 3.63^{+0.22}_{-0.35} & -1.28^{-0.52}_{+0.70}\\
 \hline\hline
 \end{array}
 \end{eqnarray*}
 \end{table*}

 \subsection{holography}
 Ref. \cite{Colangelo:2012ipa} gives almost all LECs without scalar and pseudoscalar fields in the large $N_C$ limit
 from a class of holographic theories, the results are listed in Table \ref{hol}.
 The new results still produce some large differences in these LECs, but some LECs retain the same sign,
 such as  $C_{69}$ and $C_{79}-\frac{1}{2}C_{90}$. Some differences may come from the error associated with $C_{90}$,
 the source of which needs to be further checked.

 \begin{table*}[!h]
 \extrarowheight 4pt
 \caption{The $p^6$ order LECs from Cosh holographic models \cite{Colangelo:2012ipa} and our work.
 They are in units of $10^{-3}$GeV$^{-2}$.}\label{hol}
 \begin{eqnarray*}
 \begin{array}{lccccccccc}
 \hline\hline  &C_{1}&C_{3}&C_{4}&C_{40}&C_{42}&C_{44}&C_{46}&C_{47}\\
 \hline \mbox{Ref.\cite{Colangelo:2012ipa}} & -0.3&0.3&0&0.2&2.2&-5.5&-3.2&6.2\\
 \mbox{Old\cite{our5}}& 3.79^{+0.10}_{-0.17} & -0.05^{+0.01}_{-0.01} & 3.10^{+0.09}_{-0.15} & -6.35^{-0.18}_{+0.32} & 0.60^{-0.00}_{+0.00} & 6.32^{+0.20}_{-0.36} & -0.60^{-0.02}_{+0.04} & 0.08^{+0.01}_{-0.00}\\
 \text{New}& 2.98^{+0.07}_{-0.13} & -0.05^{+0.01}_{-0.01} & 2.13^{+0.06}_{-0.10} & -4.98^{-0.14}_{+0.25} & 1.88^{+0.03}_{-0.06} & 1.73^{+0.07}_{-0.14} & -1.67^{-0.05}_{+0.09} & 3.10^{+0.09}_{-0.15}\\
 \hline\hline
 &C_{48}&C_{50}+C_{90}&C_{51}+C_{90}&C_{52}-C_{90}&C_{53}-\frac{1}{2}C_{90}&C_{55}+\frac{1}{2}C_{90}&C_{56}-C_{90}&C_{57}+2C_{90}\\
 \hline  \mbox{Ref.\cite{Colangelo:2012ipa}}&5.8&19.1&5.2&-11.6&-8.8&16.7&7.1&17.2\\
 \mbox{Old\cite{our5}}&3.41^{+0.06}_{-0.10} & 11.16^{+0.40}_{-0.66} & -9.04^{-0.20}_{+0.38} & -7.48^{-0.29}_{+0.47} & -13.21^{-0.68}_{+1.10} & 18.01^{+0.77}_{-1.26} & 16.89^{+0.89}_{-1.44} & 12.80^{+0.58}_{-0.92}\\
 \text{New}& 4.74^{+0.10}_{-0.17} & 12.86^{+0.45}_{-0.75} & -7.50^{-0.15}_{+0.30} & -8.71^{-0.33}_{+0.53} & -6.52^{-0.49}_{+0.78} & 11.61^{+0.58}_{-0.95} & 2.13^{+0.47}_{-0.73} & 9.36^{+0.48}_{-0.76}\\
 \hline\hline
 &C_{59}-\frac{1}{2}C_{90}&C_{66}&C_{69}&C_{70}-\frac{1}{2}C_{90}&C_{72}+\frac{1}{2}C_{90}&C_{73}+C_{90}&C_{74}&C_{76}-\frac{1}{2}C_{90}\\
 \hline  \mbox{Ref.\cite{Colangelo:2012ipa}}&-20.1&-0.3&0.3&5.3&-4.7&-4.4&-19.0&11.1\\
 \mbox{Old\cite{our5}}&-23.71^{-1.02}_{+1.66} & 1.71^{+0.07}_{-0.12} & -0.86^{-0.04}_{+0.06} & 0.51^{+0.11}_{-0.16} & -2.08^{-0.14}_{+0.23} & 2.94^{+0.05}_{-0.10} & -5.07^{-0.16}_{+0.27} & -2.66^{-0.05}_{+0.08}\\
 \text{New}& -15.75^{-0.79}_{+1.28} & 0.80^{+0.04}_{-0.07} & 0.52^{+0.00}_{-0.01} & 0.49^{+0.11}_{-0.16} & -0.64^{-0.10}_{+0.16} & 2.48^{+0.04}_{-0.07} & -3.34^{-0.11}_{+0.19} & -2.31^{-0.04}_{+0.07}\\
 \hline\hline
 &C_{78}+\frac{1}{2}C_{90}&C_{79}-\frac{1}{2}C_{90}&C_{87}&C_{88}-C_{90}&C_{89}\\
 \hline  \mbox{Ref.\cite{Colangelo:2012ipa}}&16.1&4.1&6.8&-5.2&29.2\\
 \mbox{Old\cite{our5}}&18.74^{+0.83}_{-1.36} & -1.78^{-0.11}_{+0.17} & 7.57^{+0.37}_{-0.60} & -7.91^{-0.35}_{+0.57} & 34.74^{+1.61}_{-2.62}\\
 \text{New}& 9.99^{+0.58}_{-0.93} & 4.70^{+0.08}_{-0.15} & 4.79^{+0.29}_{-0.46} & -4.01^{-0.24}_{+0.38} & 17.27^{+1.11}_{-1.77}\\
 \hline\hline
 \end{array}
 \end{eqnarray*}
 \end{table*}

 \subsection{Other results}
 Aside from the above results, there are more LECs given using these different method.
 Most of them are from resonance approximations, but our results do not rely on the assumption of the presence of resonances.
 In this subsection, we list the values we gathered from the literature and compare with ours.

 Table \ref{tp} lists some LECs from resonance estimates in \cite{tpf};
 Table \ref{res1} lists other resonance-estimated LECs in \cite{res2};
 Table \ref{sumr} lists the LECs from sum rules in \cite{sumrules};
 $C^r_{87}$ values were obtained from many references, and are listed Table \ref{C87};
 the other LECs obtained from different models in different references are listed in Table \ref{za}.
 Except for $C^r_{14}+C^r_{15}$ in Table \ref{za}, all the LECs retain their signs and the same orders of magnitudes,
 and almost all new results are in closer correspondence with others.

 In Table \ref{lncrr}, we checked the relations for the large $N_C$ limit given in Ref. \cite{lecr}.
 \begin{eqnarray}
 C_{20}=-3C_{21}=C_{32}=\frac{1}{6}C_{35},\hspace{1cm}C_{24}=6C_{28}=3C_{30}\;.
 \end{eqnarray}
 Because we only calculate a part of the large-$N_C$ expression in \eqref{Snorm},
 not all of the LECs satisfy the relations.

 For the anomalous parts, we collect the results in Table \ref{ap63}.
 The anomalous results are less than the normal ones,
 and the differences between each are slightly larger than the normal ones.

 \begin{table*}[!h]
 \extrarowheight 4pt
 \caption{The obtained values for the $p^6$ order LECs in Ref.\cite{tpf} and our works
 They are in units of $10^{-3}\text{ GeV}^{-2}$.}\label{tp}
 \begin{eqnarray*}
 \begin{array}{ccccccc}
 \hline\hline
 &C_{14}&C_{19}&C_{38}&C_{61}&C_{80}&C_{87}\\
 \hline
 \mbox{Ref.\cite{tpf}} & -4.3 &-2.8&1.2&1.9&1.9&7.6\\
 \mbox{Old\cite{our5}}&-0.83^{+0.12}_{-0.19}&-0.48^{+0.09}_{-0.13}&0.41^{-0.08}_{+0.07}&2.88^{-0.22}_{+0.26}&0.87^{-0.04}_{+0.03}
 &7.57^{+0.37}_{-0.60}\\
 \text{New}& -0.87^{+0.14}_{-0.21} & -0.27^{+0.09}_{-0.13} & 0.47^{-0.04}_{+0.02} & 2.42^{-0.19}_{+0.22} & 1.01^{-0.05}_{+0.04} & 4.79^{+0.29}_{-0.46}\\
 \hline\hline
 \end{array}
 \end{eqnarray*}
 \end{table*}

 \begin{table*}[!h]
 \extrarowheight 4pt
 \caption{The obtained values for the $p^6$ order LECs from resonance Lagrangian given by Ref.\cite{res2} and our work
 The coefficients in the table are in units of $10^{-4}/F_0^2$.}\label{res1}
 \begin{eqnarray*}
 \begin{array}{cccccccc}
 \hline\hline  & C_{78} & C_{82} & C_{87} & C_{88} & C_{89} & C_{90}\\
 \hline
 \mbox{Lowest Meson Dominance}   & 1.09 & -0.36 & 0.40 & -0.52 &1.97&0.0\\
 \mbox{Resonance Lagrangian I } & 1.09 & -0.29 & 0.47 & -0.16&2.29&0.33\\
 \mbox{Resonance Lagrangian II} & 1.49 & -0.39 & 0.65 & -0.14&3.22&0.51\\
 \hline
 \mbox{Old\cite{our5}} & 1.326^{+0.077}_{-0.120} & -0.539^{-0.024}_{+0.039} & 0.573^{+0.028}_{-0.045} & -0.414^{-0.055}_{+0.078}&
 2.630^{+0.122}_{-0.198}&0.185^{-0.029}_{+0.035}\\
 \mbox{New}& 0.668^{+0.060}_{-0.091} & -0.347^{-0.018}_{+0.029} & 0.363^{+0.022}_{-0.035} & -0.128^{-0.052}_{+0.070} & 1.307^{+0.084}_{-0.134} & 0.176^{-0.033}_{+0.041}\\
 \hline\hline
 \end{array}
 \end{eqnarray*}
 \end{table*}

 \begin{table*}[!h]
 \extrarowheight 4pt
 \caption{The LECs come from sum rules in \cite{sumrules}.
 They are in units of $10^{-3}\mathrm{GeV}^{-2}$.}\label{sumr}
 \begin{eqnarray*}
 \begin{array}{cccccccc}
 \hline\hline  & C_{12}+C_{61}+C_{80} & C_{12}-C_{61}+C_{80} & C_{61} & C_{12}+C_{80}\\
 \hline w_{DK}\mbox{\cite{sumrules}}& 2.48\pm0.19 & -0.55\pm0.21 & 1.51\pm0.19 & 0.97\pm0.11\\
 \hat{w}\mbox{\cite{sumrules}}& 2.48\pm0.18 & -0.46\pm0.19 & 1.47\pm0.17 & 1.01\pm0.10\\
 \mbox{Old\cite{our5}} & 3.41^{-0.25}_{+0.28} & -2.36^{+0.20}_{-0.24} & 2.88^{-0.22}_{+0.26} & 0.53^{-0.02}_{+0.02}\\
 \mbox{New}& 3.09^{-0.22}_{+0.25} & -1.75^{+0.16}_{-0.19} & 2.42^{-0.19}_{+0.22} & 0.67^{-0.03}_{+0.03}\\
 \hline\hline
 \end{array}
 \end{eqnarray*}
 \end{table*}

 \begin{table*}[!h]
 \extrarowheight 4pt
 \caption{The obtained values for the $p^6$ order LECs $C^r_{87}$.
 They are in units of $10^{-5}$.}\label{C87}
 \begin{eqnarray*}
 \begin{array}{cccccc}
 \hline\hline & \mbox{New} & \mbox{Old\cite{our5}}& \mbox{Ref.\cite{C87-1}} & \mbox{Ref.\cite{C87-2}}&\mbox{Ref.\cite{C87-3}}\\
 \hline C^r_{87} & 3.63^{+0.22}_{-0.35} & 5.73^{+0.28}_{-0.45} & 3.1\pm 1.1 & 4.3\pm 0.4 & 3.70\pm 0.14\\
 \hline\hline
 \end{array}
 \end{eqnarray*}
 \end{table*}

 \begin{table*}[!h]
 \extrarowheight 4pt
 \caption{Some LECs from different references.}\label{za}
 \begin{eqnarray*}
 \begin{array}{cccccccc}
 \hline\hline  & 10^5(2C^r_{63}-C^r_{65}) & 10^6C^r_{38} & 10^5(C^r_{88}-C^r_{90}) & (C_{14}+C_{15})10^3\text{GeV}^2& (C_{15}+2C_{17})10^3\text{GeV}^2\\
 \hline & 1.8\pm0.7\text{\cite{rkd}} & 2\pm6\text{\cite{C38-1}} & -4.6\pm0.4\text{\cite{C88C90-1}} & 0.37\pm0.08\text{\cite{latt}} &1.29\pm0.16\text{\cite{latt}} \\
                                     &&8\pm5\text{\cite{C38-2}} & -4.5\pm0.5\text{\cite{C88C90-2}}\\
 \mbox{Old\cite{our5}} & 6.36^{-0.48}_{+0.57} & 3.08^{-0.62}_{+0.56} & -5.99^{-0.27}_{+0.43}& -0.83^{+0.12}_{-0.19} & 0.03^{-0.02}_{-0.02}\\
 \mbox{New}& 5.89^{-0.45}_{+0.53} & 3.57^{-0.34}_{+0.17} & -3.04^{-0.18}_{+0.29}& -0.87^{+0.14}_{-0.21} & 0.35^{+0.03}_{-0.08}\\
 \hline\hline
 \end{array}
 \end{eqnarray*}
 \end{table*}

 \begin{table*}[!h]
 \extrarowheight 4pt
 \caption{The obtained values for the $p^6$ order LECs from our work
 The coefficients in the table are in units of $10^{-3}\mathrm{GeV}^{-2}$.}\label{lncrr}
 \begin{eqnarray*}
 \begin{array}{ccccc|ccc}
 \hline\hline  & C_{20}& -3C_{21} & C_{32}& \frac{1}{6}C_{35}&C_{24}&6C_{28}&3C_{30}\\
 \hline \mbox{Old\cite{our5}} &0.18^{-0.03}_{+0.04}&
 0.18^{-0.03}_{+0.03}&0.18^{-0.03}_{+0.04}&0.028^{-0.020}_{+0.028}&1.62^{+0.04}_{-0.07}&1.80^{+0.06}_{-0.06}&1.80^{+0.06}_{-0.09}\\
 \mbox{New}&0.17^{-0.02}_{+0.03} & 0.17^{-0.02}_{+0.03} & 0.17^{-0.02}_{+0.03} & 0.02^{-0.01}_{+0.02} & 0.87^{+0.02}_{-0.04} & 1.10^{+0.03}_{-0.05} & 1.10^{+0.03}_{-0.05}\\
 \hline\hline
 \end{array}
 \end{eqnarray*}
 \end{table*}
 \section{Summary}\label{summ}

 In this research, we updated our original LECs to the $O(p^6)$ order, including two and three flavours, and normal and anomalous ones.
 The new contributions come from $n=2$ in \eqref{Snorm} and $n=1$ in \eqref{fineqNc}.
 This is one small step beyond the GND model.
 As a check, the $O(p^4)$-order absolute values have decreased, and are closer to others,
 so our updates are plausible. Up to the $O(p^6)$ order, the absolute values of the LECs exhibited varying changes or remained unchanged.
 We also compared these LECs with others.
 Most of them are much closer than the old values, but some combinations of LECs fare badly.
 The combinations found in references are directly from phenomenological data, and they can be more precise.
 However, we have obtained LECs separately.
 On the whole, the new LECs values are better than old ones;
 one possible reason for the differences is propagation of errors.
 In this method, more precise results needs a more detailed analysis of \eqref{Snorm},
 which remains as work for the future.

 \section*{Acknowledgments}
 Jiang thanks Professor Rui-Jing Lu for the helpful discussions..
 This work was supported by the National Science Foundation of China (NSFC) under Grants No. 11205034 and No. 11475092;
 the Natural Science Foundation of Guangxi Grants No. 2013GXNSFBA019012 and No. 2013GXNSFFA019001;
  the Specialized Research Fund for the Doctoral Program of High Education of China No. 20110002110010, and the Tsinghua University Initiative Scientific Research Program No. 20121088494.

 \appendix
 \section{The $\Delta\tilde{K}_i$ coefficients}\label{tki}


 \section{The $\Delta\tilde{K}^W_i$ coefficients}\label{tobi}
 \begin{eqnarray}
 \Delta\tilde{K}^W_{1}&=&\int\frac{d^4k}{(2\pi)^4}\bigg[-\frac{1}{2} \sk^{5} X^{5}
 \bigg],\notag\\
 \Delta\tilde{K}^W_{2}&=&\int\frac{d^4k}{(2\pi)^4}\bigg[-\frac{1}{8} \sk^{3} X^{4}
 +\frac{1}{8} \sk^{5} X^{5}
 \bigg],\notag\\
 \Delta\tilde{K}^W_{3}&=&\int\frac{d^4k}{(2\pi)^4}\bigg[-\frac{1}{8} \sk^{3} X^{4}
 \bigg],\notag\\
 \Delta\tilde{K}^W_{4}&=&\int\frac{d^4k}{(2\pi)^4}\bigg[-\frac{1}{8} \sk^{3} X^{4}
 \bigg],\notag\\
 \Delta\tilde{K}^W_{5}&=&\int\frac{d^4k}{(2\pi)^4}\bigg[\frac{1}{32} \sk X^{3}
 \bigg],\notag\\
 \Delta\tilde{K}^W_{6}&=&\int\frac{d^4k}{(2\pi)^4}\bigg[\frac{1}{32} \sk^{3} X^{4}
 \bigg],\notag\\
 \Delta\tilde{K}^W_{7}&=&\int\frac{d^4k}{(2\pi)^4}\bigg[-\frac{1}{64} \sk X^{3}
 +\frac{1}{64} \sk^{3} X^{4}
 \bigg],\notag\\
 \Delta\tilde{K}^W_{8}&=&\int\frac{d^4k}{(2\pi)^4}\bigg[-\frac{1}{32} \skp^{2} X^{2}
 -\frac{11}{24} \sk\skp^{3} X^{2}
 +\frac{19}{192} X^{3}
 +\frac{1}{3} \sk\skp X^{3}
 +\frac{5}{24} \sk^{2}\skp^{2} X^{3}
 +\frac{13}{6} \sk^{3}\skp^{3} X^{3}
 \notag\\
 &&-\frac{25}{192} \sk^{2} X^{4}
 +\frac{59}{96} \sk^{4}\skp^{2} X^{4}
 -\frac{95}{24} \sk^{5}\skp^{3} X^{4}
 -\frac{1}{24} \sk^{4} X^{5}
 -\frac{43}{24} \sk^{6}\skp^{2} X^{5}
 +\frac{47}{12} \sk^{7}\skp^{3} X^{5}
 \notag\\
 &&-\frac{3}{4} \sk^{6} X^{6}
 -\frac{7}{3} \sk^{9}\skp^{3} X^{6}
 -\frac{1}{6} \sk^{8} X^{7}
 + \sk^{10}\skp^{2} X^{7}
 +\frac{2}{3} \sk^{11}\skp^{3} X^{7}
 \bigg],\notag\\
 \Delta\tilde{K}^W_{9}&=&\int\frac{d^4k}{(2\pi)^4}\bigg[\frac{1}{32} \skp^{2} X^{2}
 +\frac{5}{8} \sk\skp^{3} X^{2}
 -\frac{9}{64} X^{3}
 -\frac{1}{2} \sk\skp X^{3}
 -\frac{7}{8} \sk^{2}\skp^{2} X^{3}
 -3 \sk^{3}\skp^{3} X^{3}
 \notag\\
 &&+\frac{11}{64} \sk^{2} X^{4}
 +\frac{181}{96} \sk^{4}\skp^{2} X^{4}
 +\frac{45}{8} \sk^{5}\skp^{3} X^{4}
 +\frac{3}{8} \sk^{4} X^{5}
 -\frac{29}{24} \sk^{6}\skp^{2} X^{5}
 -\frac{23}{4} \sk^{7}\skp^{3} X^{5}
 \notag\\
 &&+\frac{13}{12} \sk^{6} X^{6}
 +\frac{5}{3} \sk^{8}\skp^{2} X^{6}
 +\frac{7}{2} \sk^{9}\skp^{3} X^{6}
 +\frac{1}{4} \sk^{8} X^{7}
 -\frac{3}{2} \sk^{10}\skp^{2} X^{7}
 - \sk^{11}\skp^{3} X^{7}
 \bigg],\notag\\
 \Delta\tilde{K}^W_{10}&=&\int\frac{d^4k}{(2\pi)^4}\bigg[\frac{7}{96} \skp^{2} X^{2}
 +\frac{5}{24} \sk\skp^{3} X^{2}
 -\frac{19}{192} X^{3}
 -\frac{1}{3} \sk\skp X^{3}
 +\frac{1}{3} \sk^{2}\skp^{2} X^{3}
 +\frac{5}{6} \sk^{3}\skp^{3} X^{3}
 \notag\\
 &&+\frac{17}{192} \sk^{2} X^{4}
 -\frac{635}{96} \sk^{4}\skp^{2} X^{4}
 -\frac{175}{24} \sk^{5}\skp^{3} X^{4}
 -\frac{5}{8} \sk^{4} X^{5}
 +\frac{143}{8} \sk^{6}\skp^{2} X^{5}
 +\frac{175}{12} \sk^{7}\skp^{3} X^{5}
 \notag\\
 &&+\frac{7}{3} \sk^{6} X^{6}
 -\frac{50}{3} \sk^{8}\skp^{2} X^{6}
 -\frac{35}{3} \sk^{9}\skp^{3} X^{6}
 -\frac{5}{6} \sk^{8} X^{7}
 +5 \sk^{10}\skp^{2} X^{7}
 +\frac{10}{3} \sk^{11}\skp^{3} X^{7}
 \bigg],\notag\\
 \Delta\tilde{K}^W_{11}&=&\int\frac{d^4k}{(2\pi)^4}\bigg[\frac{1}{24} \skp^{2} X^{2}
 +\frac{1}{3} \sk\skp^{3} X^{2}
 -\frac{13}{192} X^{3}
 -\frac{5}{24} \sk\skp X^{3}
 -\frac{19}{48} \sk^{2}\skp^{2} X^{3}
 -\frac{3}{2} \sk^{3}\skp^{3} X^{3}
 \notag\\
 &&-\frac{7}{192} \sk^{2} X^{4}
 +\frac{17}{48} \sk^{4}\skp^{2} X^{4}
 +\frac{5}{2} \sk^{5}\skp^{3} X^{4}
 +\frac{25}{48} \sk^{4} X^{5}
 +\frac{1}{2} \sk^{6}\skp^{2} X^{5}
 -\frac{13}{6} \sk^{7}\skp^{3} X^{5}
 \notag\\
 &&+\frac{1}{24} \sk^{6} X^{6}
 +\frac{7}{6} \sk^{9}\skp^{3} X^{6}
 +\frac{1}{12} \sk^{8} X^{7}
 -\frac{1}{2} \sk^{10}\skp^{2} X^{7}
 -\frac{1}{3} \sk^{11}\skp^{3} X^{7}
 \bigg],\notag\\
 \Delta\tilde{K}^W_{12}&=&\int\frac{d^4k}{(2\pi)^4}\bigg[-\frac{1}{24} \skp^{2} X^{2}
 -\frac{11}{24} \sk\skp^{3} X^{2}
 +\frac{1}{12} X^{3}
 +\frac{1}{3} \sk\skp X^{3}
 +\frac{5}{16} \sk^{2}\skp^{2} X^{3}
 +\frac{13}{6} \sk^{3}\skp^{3} X^{3}
 \notag\\
 &&-\frac{7}{24} \sk^{2} X^{4}
 +\frac{7}{16} \sk^{4}\skp^{2} X^{4}
 -\frac{95}{24} \sk^{5}\skp^{3} X^{4}
 -\frac{41}{24} \sk^{6}\skp^{2} X^{5}
 +\frac{47}{12} \sk^{7}\skp^{3} X^{5}
 -\frac{1}{2} \sk^{6} X^{6}
 \notag\\
 &&-\frac{7}{3} \sk^{9}\skp^{3} X^{6}
 -\frac{1}{6} \sk^{8} X^{7}
 + \sk^{10}\skp^{2} X^{7}
 +\frac{2}{3} \sk^{11}\skp^{3} X^{7}
 \bigg],\notag\\
 \Delta\tilde{K}^W_{13}&=&\int\frac{d^4k}{(2\pi)^4}\bigg[-\frac{1}{8} \sk^{2} X^{4}
 +\frac{1}{8} \sk^{4} X^{5}
 \bigg],\notag\\
 \Delta\tilde{K}^W_{14}&=&\int\frac{d^4k}{(2\pi)^4}\bigg[\frac{5}{96} \skp^{2} X^{2}
 +\frac{2}{3} \sk\skp^{3} X^{2}
 -\frac{23}{192} X^{3}
 -\frac{5}{12} \sk\skp X^{3}
 -\frac{31}{48} \sk^{2}\skp^{2} X^{3}
 -3 \sk^{3}\skp^{3} X^{3}
 \notag\\
 &&+\frac{1}{192} \sk^{2} X^{4}
 +\frac{49}{96} \sk^{4}\skp^{2} X^{4}
 +5 \sk^{5}\skp^{3} X^{4}
 +\frac{13}{24} \sk^{4} X^{5}
 +\frac{13}{12} \sk^{6}\skp^{2} X^{5}
 -\frac{13}{3} \sk^{7}\skp^{3} X^{5}
 \notag\\
 &&+\frac{1}{2} \sk^{6} X^{6}
 +\frac{7}{3} \sk^{9}\skp^{3} X^{6}
 +\frac{1}{6} \sk^{8} X^{7}
 - \sk^{10}\skp^{2} X^{7}
 -\frac{2}{3} \sk^{11}\skp^{3} X^{7}
 \bigg],\notag\\
 \Delta\tilde{K}^W_{15}&=&\int\frac{d^4k}{(2\pi)^4}\bigg[\frac{1}{96} \skp^{2} X^{2}
 +\frac{1}{48} \sk\skp^{3} X^{2}
 +\frac{1}{96} X^{3}
 +\frac{1}{24} \sk\skp X^{3}
 +\frac{19}{96} \sk^{2}\skp^{2} X^{3}
 -\frac{7}{12} \sk^{3}\skp^{3} X^{3}
 \notag\\
 &&+\frac{5}{48} \sk^{2} X^{4}
 -\frac{3}{16} \sk^{4}\skp^{2} X^{4}
 +\frac{125}{48} \sk^{5}\skp^{3} X^{4}
 -\frac{19}{48} \sk^{4} X^{5}
 -\frac{89}{48} \sk^{6}\skp^{2} X^{5}
 -\frac{109}{24} \sk^{7}\skp^{3} X^{5}
 \notag\\
 &&-\frac{1}{12} \sk^{6} X^{6}
 +\frac{10}{3} \sk^{8}\skp^{2} X^{6}
 +\frac{7}{2} \sk^{9}\skp^{3} X^{6}
 +\frac{1}{4} \sk^{8} X^{7}
 -\frac{3}{2} \sk^{10}\skp^{2} X^{7}
 - \sk^{11}\skp^{3} X^{7}
 \bigg],\notag\\
 \Delta\tilde{K}^W_{16}&=&\int\frac{d^4k}{(2\pi)^4}\bigg[-\frac{1}{96} \skp^{2} X^{2}
 -\frac{19}{24} \sk\skp^{3} X^{2}
 +\frac{23}{192} X^{3}
 +\frac{5}{12} \sk\skp X^{3}
 +\frac{19}{24} \sk^{2}\skp^{2} X^{3}
 +\frac{9}{2} \sk^{3}\skp^{3} X^{3}
 \notag\\
 &&-\frac{11}{64} \sk^{2} X^{4}
 -\frac{319}{96} \sk^{4}\skp^{2} X^{4}
 -\frac{85}{8} \sk^{5}\skp^{3} X^{4}
 -\frac{7}{12} \sk^{4} X^{5}
 +\frac{55}{8} \sk^{6}\skp^{2} X^{5}
 +\frac{163}{12} \sk^{7}\skp^{3} X^{5}
 \notag\\
 &&+\frac{1}{12} \sk^{6} X^{6}
 -\frac{25}{3} \sk^{8}\skp^{2} X^{6}
 -\frac{28}{3} \sk^{9}\skp^{3} X^{6}
 -\frac{2}{3} \sk^{8} X^{7}
 +4 \sk^{10}\skp^{2} X^{7}
 +\frac{8}{3} \sk^{11}\skp^{3} X^{7}
 \bigg],\notag\\
 \Delta\tilde{K}^W_{17}&=&\int\frac{d^4k}{(2\pi)^4}\bigg[\frac{1}{24} \skp^{2} X^{2}
 -\frac{17}{24} \sk\skp^{3} X^{2}
 +\frac{1}{12} X^{3}
 +\frac{1}{3} \sk\skp X^{3}
 -\frac{1}{16} \sk^{2}\skp^{2} X^{3}
 +\frac{31}{6} \sk^{3}\skp^{3} X^{3}
 \notag\\
 &&-\frac{1}{6} \sk^{2} X^{4}
 -\frac{35}{16} \sk^{4}\skp^{2} X^{4}
 -\frac{365}{24} \sk^{5}\skp^{3} X^{4}
 +\frac{245}{24} \sk^{6}\skp^{2} X^{5}
 +\frac{269}{12} \sk^{7}\skp^{3} X^{5}
 +\frac{1}{4} \sk^{6} X^{6}
 \notag\\
 &&-15 \sk^{8}\skp^{2} X^{6}
 -\frac{49}{3} \sk^{9}\skp^{3} X^{6}
 -\frac{7}{6} \sk^{8} X^{7}
 +7 \sk^{10}\skp^{2} X^{7}
 +\frac{14}{3} \sk^{11}\skp^{3} X^{7}
 \bigg],\notag\\
 \Delta\tilde{K}^W_{18}&=&\int\frac{d^4k}{(2\pi)^4}\bigg[\frac{1}{12} \skp^{2} X^{2}
 +\frac{2}{3} \sk\skp^{3} X^{2}
 -\frac{13}{96} X^{3}
 -\frac{5}{12} \sk\skp X^{3}
 -\frac{19}{24} \sk^{2}\skp^{2} X^{3}
 -3 \sk^{3}\skp^{3} X^{3}
 \notag\\
 &&-\frac{1}{32} \sk^{2} X^{4}
 +\frac{17}{24} \sk^{4}\skp^{2} X^{4}
 +5 \sk^{5}\skp^{3} X^{4}
 +\frac{7}{12} \sk^{4} X^{5}
 + \sk^{6}\skp^{2} X^{5}
 -\frac{13}{3} \sk^{7}\skp^{3} X^{5}
 \notag\\
 &&+\frac{1}{2} \sk^{6} X^{6}
 +\frac{7}{3} \sk^{9}\skp^{3} X^{6}
 +\frac{1}{6} \sk^{8} X^{7}
 - \sk^{10}\skp^{2} X^{7}
 -\frac{2}{3} \sk^{11}\skp^{3} X^{7}
 \bigg],\notag\\
 \Delta\tilde{K}^W_{19}&=&\int\frac{d^4k}{(2\pi)^4}\bigg[\frac{1}{96} \skp^{2} X^{2}
 -\frac{1}{4} \sk\skp^{3} X^{2}
 +\frac{1}{64} X^{3}
 +\frac{11}{16} \sk^{2}\skp^{2} X^{3}
 +3 \sk^{3}\skp^{3} X^{3}
 -\frac{3}{64} \sk^{2} X^{4}
 \notag\\
 &&-\frac{595}{96} \sk^{4}\skp^{2} X^{4}
 -\frac{45}{4} \sk^{5}\skp^{3} X^{4}
 -\frac{1}{4} \sk^{4} X^{5}
 +\frac{97}{6} \sk^{6}\skp^{2} X^{5}
 +\frac{37}{2} \sk^{7}\skp^{3} X^{5}
 +\frac{7}{6} \sk^{6} X^{6}
 \notag\\
 &&-\frac{50}{3} \sk^{8}\skp^{2} X^{6}
 -14 \sk^{9}\skp^{3} X^{6}
 - \sk^{8} X^{7}
 +6 \sk^{10}\skp^{2} X^{7}
 +4 \sk^{11}\skp^{3} X^{7}
 \bigg],\notag\\
 \Delta\tilde{K}^W_{20}&=&\int\frac{d^4k}{(2\pi)^4}\bigg[\frac{1}{24} \skp^{2} X^{2}
 +\frac{5}{12} \sk\skp^{3} X^{2}
 -\frac{1}{24} X^{3}
 -\frac{1}{6} \sk\skp X^{3}
 -\frac{7}{8} \sk^{2}\skp^{2} X^{3}
 -\frac{5}{3} \sk^{3}\skp^{3} X^{3}
 \notag\\
 &&+\frac{1}{8} \sk^{2} X^{4}
 +\frac{9}{4} \sk^{4}\skp^{2} X^{4}
 +\frac{25}{12} \sk^{5}\skp^{3} X^{4}
 +\frac{1}{4} \sk^{4} X^{5}
 -\frac{17}{12} \sk^{6}\skp^{2} X^{5}
 -\frac{5}{6} \sk^{7}\skp^{3} X^{5}
 \bigg],\notag\\
 \Delta\tilde{K}^W_{21}&=&\int\frac{d^4k}{(2\pi)^4}\bigg[-\frac{1}{4} \sk\skp^{3} X^{2}
 +\frac{19}{24} \sk^{2}\skp^{2} X^{3}
 +3 \sk^{3}\skp^{3} X^{3}
 +\frac{1}{24} \sk^{2} X^{4}
 -\frac{51}{8} \sk^{4}\skp^{2} X^{4}
 -\frac{45}{4} \sk^{5}\skp^{3} X^{4}
 \notag\\
 &&-\frac{1}{3} \sk^{4} X^{5}
 +\frac{65}{4} \sk^{6}\skp^{2} X^{5}
 +\frac{37}{2} \sk^{7}\skp^{3} X^{5}
 +\frac{7}{6} \sk^{6} X^{6}
 -\frac{50}{3} \sk^{8}\skp^{2} X^{6}
 -14 \sk^{9}\skp^{3} X^{6}
 \notag\\
 &&- \sk^{8} X^{7}
 +6 \sk^{10}\skp^{2} X^{7}
 +4 \sk^{11}\skp^{3} X^{7}
 \bigg],\notag\\
 \Delta\tilde{K}^W_{22}&=&\int\frac{d^4k}{(2\pi)^4}\bigg[\frac{1}{16} \skp^{2} X^{2}
 +\frac{1}{16} \sk\skp^{3} X^{2}
 -\frac{13}{32} \sk^{2}\skp^{2} X^{3}
 -\frac{1}{4} \sk^{3}\skp^{3} X^{3}
 -\frac{1}{16} \sk^{2} X^{4}
 +\frac{21}{32} \sk^{4}\skp^{2} X^{4}
 \notag\\
 &&+\frac{5}{16} \sk^{5}\skp^{3} X^{4}
 +\frac{1}{16} \sk^{4} X^{5}
 -\frac{5}{16} \sk^{6}\skp^{2} X^{5}
 -\frac{1}{8} \sk^{7}\skp^{3} X^{5}
 \bigg],\notag\\
 \Delta\tilde{K}^W_{23}&=&\int\frac{d^4k}{(2\pi)^4}\bigg[-\frac{1}{8} \sk\skp^{3} X^{2}
 +\frac{9}{16} \sk^{2}\skp^{2} X^{3}
 +\frac{3}{2} \sk^{3}\skp^{3} X^{3}
 -\frac{63}{16} \sk^{4}\skp^{2} X^{4}
 -\frac{45}{8} \sk^{5}\skp^{3} X^{4}
 -\frac{1}{4} \sk^{4} X^{5}
 \notag\\
 &&+\frac{73}{8} \sk^{6}\skp^{2} X^{5}
 +\frac{37}{4} \sk^{7}\skp^{3} X^{5}
 +\frac{11}{16} \sk^{6} X^{6}
 -\frac{35}{4} \sk^{8}\skp^{2} X^{6}
 -7 \sk^{9}\skp^{3} X^{6}
 -\frac{1}{2} \sk^{8} X^{7}
 \notag\\
 &&+3 \sk^{10}\skp^{2} X^{7}
 +2 \sk^{11}\skp^{3} X^{7}
 \bigg].\label{dkwt}
 \end{eqnarray}

 \bibliography{getlecs}
\end{document}